\documentclass[pra,twocolumn,aps,amssymb,showpacs,superscriptaddress]{revtex4-1}

\usepackage{epsfig}
\usepackage{graphicx}
\usepackage{amsmath}
\usepackage{amssymb}
\usepackage{amsmath,amssymb}
\usepackage{enumerate}
\usepackage{hyperref}
\usepackage[normalem]{ulem}
\usepackage{cancel}
\usepackage{float}
\usepackage{comment}
\usepackage{bm}
\usepackage{color}
\definecolor{rosso}{rgb}{1,0,0}
\definecolor{verde}{rgb}{0,1,0}
\definecolor{blue}{rgb}{0,0,1}
\definecolor{verdescuro}{rgb}{0,0.5,0.5}
\definecolor{rossoscuro}{rgb}{0.7,0.3,0}
\definecolor{bluscuro}{rgb}{0.3,0,0.7}
\definecolor{magenta}{rgb}{1,0,1}

\renewcommand{\Re}{\operatorname{Re}}
\renewcommand{\Im}{\operatorname{Im}}
\newcommand{\om}{{\omega}}
\newcommand{\Om}{{\Omega}}
\newcommand{\Omp}{{\Omega_{\rm p}}}
\newcommand{\omth}{\varepsilon_0^*}
\newcommand{\Omth}{{\Omega_{\rm th}}}
\newcommand{\omrf}{{\omega_{\delta}}}

\newcommand{\Irf}{{I_{\rm rf}}}
\newcommand{\vQ}{{\bf Q}}

\newcommand{\vk}{{\bf k}}
\newcommand{\vp}{{\bf p}}

\newcommand{\xik}{{\xi_{k}}}
\newcommand{\xip}{{\xi_{p}}}
\newcommand{\be}{{\varepsilon_{0}}}
\newcommand{\coup}{{(k_{\rm F}a_{\rm F})^{-1}}}

\newcommand{\tk}{{\tilde{k}}}

\newcommand\beq{\begin{equation}}
\newcommand\eeq{\end{equation}}
\newcommand{\bea}{\begin{eqnarray}}
\newcommand{\eea}{\end{eqnarray}}

\begin{document}

\title{Peaks and widths of radio-frequency spectra: \\ An analysis of the phase diagram of ultra-cold Fermi gases}
\author{Leonardo Pisani}
\affiliation{School of Science and Technology, Physics Division, Universit\`{a} di Camerino, 62032 Camerino, Italy}
\author{Michele Pini}
\affiliation{Max Planck Institute for the Physics of Complex Systems, N\"{o}thnitzer Str. 38, 01187 Dresden, Germany}
\author{Pierbiagio Pieri}
\affiliation{Dipartimento di Fisica e Astronomia, Universit\`a di Bologna, Via Irnerio 46, 40126 Bologna, Italy}
\affiliation{INFN, Sezione di Bologna, Viale Berti Pichat 6/2, 40127 Bologna, Italy}
\author{Giancarlo Calvanese Strinati}
\affiliation{School of Science and Technology, Physics Division, Universit\`{a} di Camerino, 62032 Camerino, Italy}
\affiliation{CNR-INO, Istituto Nazionale di Ottica, Sede di Firenze, 50125 Firenze, Italy}

\date{\today}

\begin{abstract}
We provide a comprehensive theoretical study of the radio-frequency (rf) spectra of a two-component Fermi gas with balanced populations in the normal region of the temperature-vs-coupling phase diagram. 
In particular, rf spectra are analyzed in terms of two characteristic peaks, which can be either distinct or overlapping. 
On the BEC side of the crossover, these two contributions are associated with a fermionic quasi-particle peak and a bosonic-like contribution due to pairing. 
On the BCS side of the crossover, the two peaks are  instead associated with interactions between particles occurring, respectively, at high or low relative momenta.    
Through this two-peak analysis, we show how and to what extent the correlation between the widths of the rf spectra and the pair size, previously identified in the superfluid phase at low temperature, can be extended to the normal phase, 
as well as how the temperature-vs-coupling phase diagram of the BCS-BEC crossover can be partitioned in a number of distinct physical sectors. 
Several analytic results for the shape and widths of the rf spectra are also derived in appropriate temperature and coupling limits.
\end{abstract}

\maketitle

\section{Introduction} 
\label{sec:introduction}

Radio-frequency (rf) spectroscopy is widely used in experiment with ultra-cold Fermi gases, both as a tool for their preparation (for instance, to adjust the populations of hyperfine levels), or as a probe of their physical properties~\cite{Ketterle-2008,Vale-2021}.   
With rf spectroscopy, mean-field shifts and scattering lengths were measured in Refs.~\cite{Regal-2003-a,Gupta-2003} across the Fano-Feshbach resonance that drives the BCS-BEC crossover, while in Ref.~\cite{Regal-2003-b}
the binding energy of the molecules formed on the BEC side of the crossover was measured.

From a many-body physics perspective, a surge of interest was initially motivated by the work of Ref.~\cite{Chin-2004}, where rf spectroscopy was used with the aim of measuring the pairing gap across the BCS-BEC crossover. 
Subsequently, several experimental papers  have appeared on the subject \cite{Shin-2007,Stewart-2008,Schirotzek-2008,Schunck-2008,Gaebler-2010,Shkedrov-2018,Mukherjee-2019}, 
accompanied by a number of theoretical papers which contributed to the  physical interpretations of rf spectra \cite{Kinnunen-2004,He-2005,Ohashi-2005,Yu-2006,Punk-2007,Baym-2007,Perali-2008,Massignan-2008,Pieri-2009,Haussmann-2009,Pieri-2011,Torma-2016}.  
The conclusion of these experimental and theoretical studies about the original aim of Ref.~\cite{Chin-2004} was that extracting the pairing gap from the rf spectra is actually hindered by several concurring effects 
which contribute to shaping the rf spectra. 

From a different perspective, Ref.~\cite{Schunck-2008} has showed that what can be extracted from the shape of rf spectra is the Cooper pair size at low temperature along the BCS-BEC crossover \cite{Pistolesi-1994,Pistolesi-1996,Palestini-2014}.
In addition, a recent experimental work \cite{Mukherjee-2019} has reported rf spectra of the unitary Fermi gas in a uniform box over a wide temperature range (up to several times the Fermi temperature $T_{\rm F}$), showing a gradual evolution from quantum to classical physics. 

Here, we provide a comprehensive theoretical study of the rf spectra of a two-component Fermi gas with balanced populations in the normal region of the temperature-vs-coupling phase diagram. 
In particular, we are able to interpret the rf spectra in terms of two characteristic peaks, which can be either distinct or overlapping (or, sometimes, even completely covering each other).
On the BEC side of the crossover, the two peaks are associated with a bosonic-like contribution due to pairing and with a fermionic quasi-particle peak. 
This sharp characterization of the two peaks evolves as one moves to the BCS side of the crossover. 
More generally, the two peaks are associated with contributions to the single-particle self-energy resulting from low- and high-energy scattering processes, respectively. 
 
Through this two-peak analysis, we show that the correlation between the widths of the rf spectra and the pair size, that was previously identified in the superfluid phase at low temperature \cite{Schunck-2008}, 
persists also in the normal phase in an appropriate region of the phase diagram.
More generally, we exploit the two-peak analysis to partition the normal region of the temperature-vs-coupling phase diagram of the BCS-BEC crossover into a number of distinct sectors, where the two-component Fermi gas effectively behaves in different ways.
In addition, through this analysis we obtain several useful analytic results about the shape and width of rf spectra in appropriate temperature and coupling regions.

Our calculations of rf spectra are based on a diagrammatic $t$-matrix approach, which has previously been validated against experimental rf spectroscopy data in both three~\cite{Gaebler-2010,Tsuchiya-2010,Perali-2011,Ota-2017,Hu-2022} and two~\cite{Schmidt-2012,Pietila-2012,Watanabe-2013,Marsiglio-2015} dimensions.

The article is organized as follows.
Section~\ref{sec:theoretical_approach} describes the theoretical approach adopted to obtain the rf spectra and the two-peak analysis that we have implemented to interpret the results. 
Section~\ref{sec:phdiag} describes the ensuing partitioning of the temperature-vs-coupling phase diagram in different sectors and discusses the physical features of each sector (as summarized in Fig.~\ref{Figure-11} therein). 
Section~\ref{sec:conclusions} gives our conclusions. 
The Appendices report several results of analytic calculations, which are of help in extracting the relevant physical features from the rf spectra.
Appendix~\ref{appA} discusses how the bound-state contribution to the self-energy is computed and provides its analytical expression in the BEC limit.
Appendix~\ref{appB} derives the momentum-resolved rf spectra for a thermal assembly of non-interacting molecules (thus generalizing the experimental fitting function introduced in Ref.~\cite{Sagi-2015}),
and compares it favorably with the low- and high-temperature BEC limits of the momentum-resolved rf spectra obtained by our $t$-matrix approach.
Appendix \ref{appCwhole} provides analytical expressions for the rf spectra at high-temperature in the BEC limit.
An asymptotic expression is also found for the corresponding full width at half maximum (FWHM) in an appropriate temperature regime.
Appendix~\ref{appD} derives the FWHM of the rf spectra in the Boltzamnn limit at unitarity and in the weak-coupling (BCS) limit. 
Throughout, the reduced Planck constant $\hbar$ and the Boltzmann constant $k_{\rm B}$ are set equal to unity for convenience.

\section{Rf spectra within a many-body approach} 
\label{sec:theoretical_approach}

In this Section, we describe the theoretical many-body scheme that we have adopted for calculating the rf spectra in the normal region of the temperature-vs-coupling phase diagram.
In addition, from an analysis of the rf spectra obtained in this way at several temperatures for given coupling, we identify the presence of a ``fixed point'' in the spectra that reveals the presence of two
underlying peaks.
We also show how important physical information can be extracted from this two-peak analysis.
We emphasize that our analysis is performed for a \emph{homogeneous} Fermi gas, and is thus relevant for experiments using either a box-like trapping potential \cite{Mukherjee-2017,Hueck-2018,Yan-2019,Mukherjee-2019,Shkedrov-2022} or tomographic reconstruction of the homogeneous rf signal from the original trap-averaged one \cite{Shin-2007,Schirotzek-2008,Schunck-2008}.
\vspace{-0.3cm}
\subsection{Theoretical formalism for rf spectroscopy in the normal phase of a balanced Fermi gas}
\label{sec:nsctheory}

We consider a two-component atomic Fermi gas, corresponding to a balanced mixture of atoms of the same species of mass $m$ in two different hyperfine levels (labeled by an effective spin variable $\sigma=\uparrow,\downarrow$) and mutually interacting through a contact interaction, as described by the Hamiltonian:
 \begin{eqnarray}
\hat{H}&=& \sum_{\sigma}\int d {\bf r} \, \hat{\psi}^{\dagger}_{\sigma}({\bf r}) \left( -\frac{\nabla^2}{2m} \right) \hat{\psi}_{\sigma}({\bf r}) \nonumber \\
&+& v_0\int d {\bf r} \, \hat{\psi}^{\dagger}_{\uparrow}({\bf r}) \hat{\psi}^{\dagger}_{\downarrow}({\bf r}) \hat{\psi}_{\downarrow}({\bf r}) \hat{\psi}_{\uparrow}({\bf r}) \, .
\label{H}
 \end{eqnarray}
Here, $\hat{\psi}_{\sigma}({\bf r})$ is a field operator with spin projection $\sigma$ and $v_0$ is the bare interaction strength ($v_0\to0^{-}$ when the contact interaction is regularized in terms of the two-fermion scattering length 
$a_{\rm F}$~\cite{Sademelo-1993,Pieri-2000}).

This model was first considered to describe the BCS-BEC crossover in a 3D attractive Fermi gas in Ref.~\onlinecite{Sademelo-1993} and then 
used extensively in the literature on the BCS-BEC crossover.
In particular, in the context of ultra-cold Fermi gases this model can be justified from first principles as an effective model describing the system when the interaction is modulated by a broad Fano-Feshbach resonance \cite{Simonucci-2005}. 
 
\subsubsection{Radio-frequency spectral intensity and single-particle spectral function}

In rf spectroscopy, a radio-frequency pulse is used to transfer atoms from one of the two hyperfine level (say, level $\downarrow$) to a third final state $f$.  
For given detuning frequency $\omega_{\delta} = \omega_{\rm rf} - \omega_{\downarrow f}$ of the rf field $\omega_{\rm rf}$ from the bare atomic transition frequency $\omega_{\downarrow f}$, the experiment measures the number of atoms per unit time $N_{\rm rf}(\omega_{\delta})$ transferred from $\vert\!\downarrow\rangle$ to $\vert f \rangle$. 
Under the standard experimental conditions of a weak rf field and of a small detuning frequency $\omega_\delta$ compared with the bare atomic transition frequency $\omega_{\downarrow f}$, the rotating-wave approximation is applicable 
such that the interaction Hamiltonian with the rf electromagnetic wave with magnetic field component ${\bf B}_{\rm rf}(t)={\bf B}_{\rm rf}^0 \cos(\omega_{\rm rf} t)$ reads 
\begin{equation}
\hat{H}'(t)= \frac{\Omega_{R}}{2}\int d {\bf r} \, e^{-i\omega_{\rm rf}t}\hat{\psi}_{\rm f}^{\dagger}({\bf r})\hat{\psi}_{\downarrow}({\bf r}) + {\rm h.c.} \, ,
\end{equation}
where $\Omega_{\rm R}= |\langle {\rm f} \vert  {\bm \mu}\cdot {\bf B}_{\rm rf}^0 \vert \!\downarrow \rangle |$ is the Rabi frequency and ${\bm \mu}$ the total magnetic moment operator of the atom. 

The assumption of weak rf field justifies also the use of linear-response theory to calculate $N_{\rm rf}(\omega_{\delta})$, which is thus given by
~\cite{He-2005,Punk-2007,Massignan-2008,Pieri-2009}
 \begin{equation}
N_{\rm rf}(\omega) = - \frac{\Omega_{\rm R}^2}{2} \int \!\! d{\bf r} \int \!\! d {\bf r}' \, {\rm Im} \Pi^{\rm R}_{\downarrow {\rm f}}({\bf r},{\bf r}',\omega+\mu_{\downarrow}-\mu_{\rm f}+\omega_{\downarrow {\rm f}}) 
\label{rf-spectrum}
\end{equation}
where $\Pi^{\rm R}_{\downarrow {\rm f}}({\bf r},{\bf r}',\omega')=\int dt  \, e^{i\omega't}\Pi^{\rm R}_{\downarrow {\rm f}}({\bf r},{\bf r}',t)$ is the Fourier transform at frequency $\omega'$ of the retarded correlation function
\begin{equation}
\Pi^{\rm R}_{\downarrow {\rm f}}({\bf r},{\bf r}',t)= 
-i \theta(t)\langle[\hat{\psi}_{\downarrow}^{\dagger}({\bf r},t)
\hat{\psi}_{\rm f}({\bf r},t),\hat{\psi}^{\dagger}_{\rm f}({\bf r}',0)\hat{\psi}_{\downarrow}({\bf r}',0]\rangle .
\label{Pi_r}
\end{equation}
Note that the time evolution of the field operators in Eq.~(\ref{Pi_r}) is determined by the grand-canonical Hamiltonian $\hat{K}=\hat{H}-\sum_s \hat{N}_s \mu_s$   (where $s=\downarrow,\uparrow,$f).
The rf spectrum (\ref{rf-spectrum}) obeys the sum rule~\cite{Perali-2008}
\begin{equation}
\int_{-\infty}^{+\infty} \!\!\!\! N_{\rm rf}(\omega) d \omega=\frac{\Omega_{\rm R}^2}{2} \pi(N_{\downarrow}-N_{\rm f}) \, ,
\label{sumrule}
\end{equation}
where $N_{\downarrow}$ and $N_{\rm f}$  are the total numbers of atoms in the states $|\!\downarrow\rangle$ and $\vert {\rm {\rm f}} \rangle$ at equilibrium (i.e., before the rf field is applied). 
For the two-component balanced mixture of interest here, $N_{\downarrow}=N_{\uparrow}=N/2$ and $N_{\rm f}=0$.

For a uniform system confined in a volume $V$, in the  absence of final-state interaction between the state $f$ and the states $(\uparrow,\downarrow)$, the spectral intensity reduces to~\cite{Kinnunen-2004,He-2005,Punk-2007,Perali-2011}
\begin{equation}
N_{\rm rf}(\omega_\delta) =  \frac{V\Omega_{\rm R}^2\pi}{2}\int \!\frac{d{\bf k}}{(2\pi)^3} A({\bf k},\xik-\omega_\delta)f(\xik-\omega_\delta) \, ,
\label{N_omega}
\end{equation}   
where $\xik=k^2/(2m)-\mu$ with chemical potential $\mu$, $f(x)=[\exp(x/T)+1]^{-1}$ is the Fermi function at temperature $T$, and $A({\bf k},\omega)$ is the single-particle spectral function at wave vector ${\bf k}$ and frequency $\omega$ 
which is determined by the retarded Green's function $G^{\rm R}({\bf k},\omega)$:
\begin{equation}
A({\bf k},\omega)=-\frac{1}{\pi} {\rm Im} G^{\rm R}({\bf k},\omega) \, .
\end{equation}

In the following, it is convenient to introduce dimensionless quantities by expressing energies (and frequencies) in units of the Fermi energy $E_{\rm F}$ and wave vectors in units of $k_{\rm F}$, where 
$E_{\rm F} = k_{\rm F}^2/(2 m)$ and $k_{\rm F}=(3\pi^2 n)^{1/3}$, $n=N/V$ being the particle number density. 
One may thus define the dimensionless spectral intensity 
\begin{equation}
I_{\rm rf}(\bar{\omega}_{\delta}) = \frac{2 E_{\rm F}}{\pi N_{\downarrow}\Omega_{\rm R}^2} N_{\rm rf}(\bar{\omega}_\delta) \, ,
\end{equation}
such that the sum rule (\ref{sumrule}) reads 
\begin{equation}
\int_{-\infty}^\infty \!\!\! d \bar{\omega_\delta} \, I_{\rm rf}(\bar{\omega}_{\delta}) = 1 ,
\end{equation}
while Eq.~(\ref{N_omega}) becomes 
\begin{equation}
I_{\rm rf}(\bar{\omega}_{\delta}) = 3 \int_0^{\infty} \!\!\! d\bar{k} \, \bar{k}^2 E_{\rm F} A(\bar{{\bf k}},\bar{\xi}_{\bar{k}}-\bar{\omega}_\delta)f(\bar{\xi}_{\bar{k}}-\bar{\omega}_\delta) \, .
\label{irfnorm}
\end{equation}
In the above expressions, the rotational invariance of the Hamiltonian (\ref{H}) has been exploited and the over-line indicates dimensionless quantities.

\vspace{-0.3cm}
\subsubsection{Choice of the single-particle self-energy}

Theoretical calculations of the single-particle spectral function $A({\bf k},\omega)$ require a specific choice of the single-particle self-energy.  
As mentioned in the Introduction, in the following we shall adopt the (non-self-consistent) $t$-matrix approximation \cite{Haussmann-1993,Haussmann-1994,Micnas-1995,Yanase-1999,Rohe-2001,Perali-2002} for the self-energy, which has previously been validated against experimental rf spectroscopy data in both three~\cite{Gaebler-2010,Tsuchiya-2010,Perali-2011,Ota-2017,Hu-2022} and two~\cite{Schmidt-2012,Pietila-2012,Watanabe-2013,Marsiglio-2015} dimensions.
This choice of the self-energy takes into account the effects of pairing fluctuations in the normal phase of an attractive Fermi gas, and is given by the following expression \cite{Yanase-1999,Rohe-2001}:
\beq
\Sigma(\vk,\om_n)=\int \frac{d\vQ}{(2\pi)^3}  \, T  \sum_\nu \Gamma_0(\vQ,\Om_\nu)G_0(\vQ-\vk,\Om_\nu-\om_n) \, . 
\label{selen}
\eeq
Here, $\om_n=(2n+1)\pi T$ and  $\Om_\nu=2\pi\nu T$ ($n,\nu$ integer) are fermionic and bosonic Matsubara frequencies, respectively,  
$G_0(\vk,\om_n)=(i\om_n-\xik)^{-1}$ is the bare fermionic single-particle propagator, and 
\bea
&&\Gamma_0({\bf Q},\Omega_{\nu})=-\left\{\frac{m}{4\pi a_{\rm F}} + \int\frac{d{\bf k}}{(2\pi)^3}\right.\nonumber\\
&&\left.\times \left[T\sum_n G_0({\bf k},\omega_n)G_0({\bf Q}-{\bf k},\Omega_{\nu}-\omega_n)-\frac{m}{k^2}\right]\right\}^{-1}
\label{gamma0}
\eea
is the pair (or particle-particle) propagator. 
Note that in Eq.~(\ref{gamma0}) we have used the regularization of the contact interaction in terms of the scattering length $a_{\rm F}$ \cite{Sademelo-1993,Pieri-2000}.

Analytic continuation of the self-energy~(\ref{selen}) from Matsubara to real frequencies is required to obtain dynamical quantities, like the the single-particle spectral function $A(\vk,\omega)$ of interest.
This is achieved  by introducing the spectral representation of the pair propagator~\cite{Yanase-1999,Rohe-2001,Pieri-2004}
\beq
\Gamma_0(\vQ,\Om_\nu)=-\int_{-\infty}^{\infty} \frac{d\Om}{\pi} \frac{\Im\Gamma_{0}^R(\vQ,\Om)}{i\Om_\nu-\Om} \, ,
\label{img0R}
\eeq
where the pair spectral function $\Im\Gamma_{0}^R(\vQ,\Om)$ is, in turn, obtained by analytic continuation of the right-hand side of Eq.~(\ref{gamma0}) with the replacement $i\Om_\nu \rightarrow \Om+i 0^+$. 

By entering Eq.~(\ref{img0R}) into Eq.~(\ref{selen}), performing the sum over the Matsubara frequency $\Om_\nu$, and then letting $i\om_n \rightarrow \om+i 0^+$, one obtains the following expression for
the imaginary part of the self-energy on the real frequency axis \cite{Yanase-1999,Rohe-2001,Perali-2002}
\begin{align}
\Im\left[\Sigma(\mathbf{k}, \om)\right]&= -\int \frac{d \vQ}{(2 \pi)^{3}}  \Im\left[\Gamma_{0}^R\left(\vQ, \omega+\xi_{\vQ-\vk}\right)\right] \nonumber \\ 
                              & \times  \left[b\left(\omega+\xi_{\vQ-\vk}\right)+f\left(\xi_{\vQ-\vk}\right)\right] \, . 
\label{imsancon}
\end{align}
where $b(\omega)$ and $f(\omega)$ are the Bose and Fermi distribution functions, respectively.
The real part of $\Sigma$ is then obtained via a Kramers-Kronig transform. 
The self-energy that results in this way is then inserted into the Dyson's equation $G^{-1}=G_0^{-1} -\Sigma$ to yield the single-particle spectral function, which in terms  of the real and imaginary parts of the self-energy reads:
\beq
A({\bf k},\omega)=--\frac{1}{\pi}\frac{\Im \Sigma(\vk,\om)}{[\om-\xik-\Re \Sigma(\vk,\om)]^2 + [\Im \Sigma(\vk,\om)]^2} \, .
\label{akwdef}
\eeq
For a given choice of temperature, density, and scattering length, the chemical potential
$\mu$ is determined by inverting the density equation
\beq
 \frac{n}{2} = \int \!\! \frac{d \vk}{(2 \pi)^{3}} \int \!\!d\om f(\om)  A(\vk,\om) \label{densequ} \, ,
\eeq
or by its equivalent version in Matsubara space
\beq
 \frac{n}{2} = \int \!\! \frac{d \vk}{(2 \pi)^{3}} \,T  \sum_n G(\vk,\omega_n) e^{i\omega_n 0^+}\label{densequM}.
\eeq
Comparison between the results obtained by using alternatively Eq.~(\ref{densequ}) or Eq. (\ref{densequM}) provides us with a check on the numerical calculations. 

It should be mentioned that the present non-self-consistent $t$-matrix approach, when the Dyson’s equation is expanded to first order, coincides with the diagrammatic approach introduced by Nozi\`eres and Schmitt-Rink in their foundational work on the BCS-BEC crossover~\cite{Nozieres-1985} (as first noticed in Ref.~\onlinecite{Serene-1989}). 
It thus shares with Ref.~\onlinecite{Nozieres-1985} the same correct behavior for the superfluid critical temperature obtained in terms of the Thouless criterion \cite{Thouless-1960}, $\Gamma_0(0,0)^{-1}=0$), and accordingly recovers the expected BCS and BEC transition temperatures in the two opposite limits of the BCS-BEC crossover.
In these limits, the self-energy becomes small compared to the chemical potential $\mu$, thus justifying the expansion of the Dyson's equation and making the two (non-self-consistent $t$-matrix and Nozi\`eres and Schmitt-Rink) approaches equivalent to each other \cite{Physics-Reports-2018}. 
In addition, as first pointed out in Ref.~\cite{Combescot-2006},  for all coupling strengths the $t$-matrix approach recovers  in the high-temperature limit the leading contribution due to interaction in the virial expansion \cite{Liu-2009,Leyronas-2011,Liu-2013} and is thus correct to second order in the small parameter provided by the fugacity $z= e^{\beta \mu}$.
\begin{figure*}[t]
\begin{center}
\includegraphics[width=\textwidth]{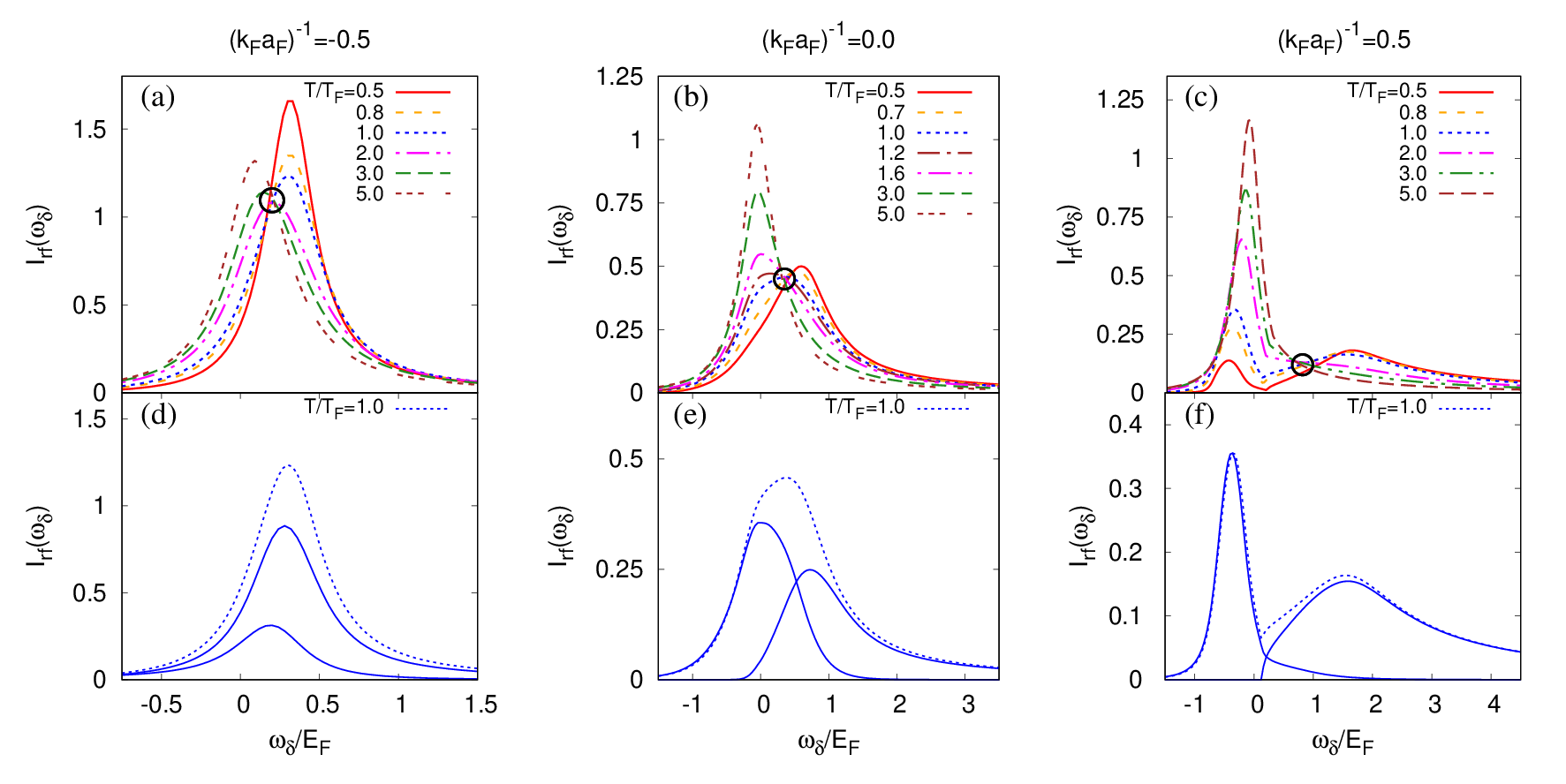}
\caption{Top panels: Dimensionless rf spectral intensity $I_{\rm rf}$ as a function of detuning frequency $\omega_{\delta}$ (in units of $E_{\rm F}$), for different temperatures and three representative couplings across the BCS-BEC crossover.
                                 The meaning of the empty circles is explained in the text.
              Bottom panels: Two-component fit of the spectra in the corresponding top panels for the representative case with $T=T_{\rm F}$.} 
\label{Figure-1}
\end{center}
\end{figure*} 

Note that, when physical quantities are made dimensionless according to the procedure described above, all quantities depend just on $T/T_{\rm F}$ and the dimensionless coupling parameter $(k_{\rm F} a_{\rm F})^{-1}$, which drives the crossover from the BCS to BEC regimes, corresponding to $(k_{\rm F} a_{\rm F})^{-1}\lesssim -1$ and $(k_{\rm F} a_{\rm F})^{-1}\gtrsim +1$, respectively. 
The so-called unitarity limit, where the scattering length diverges and  $(k_{\rm F} a_{\rm F})^{-1}=0$, then corresponds to the middle of the crossover region in between the BCS and BEC regimes.

Finally, we recall that alternative $t$-matrix approaches have been considered in the literature, corresponding to different level of self-consistency in the self-energy (see Ref.~\cite{Pini-2019} for a comprehensive discussion of the different schemes), at the extremes of which are the non-self-consistent $t$-matrix approach and the fully self-consistent one. 
The fully self-consistent approach compares better with experimental and Quantum Monte Carlo results for \emph{static\/} thermodynamic quantities~\cite{Ku-2012,Zwerger-2016,Carcy-2019,Mukherjee-2019,Jensen-2020,Rammelmueller-2021}. 
On the other hand, it is known from many-body theory of condensed-matter systems that for \emph{dynamic\/} quantities like the spectral weight function, the inclusion of self-consistency in the Green's function without the simultaneous inclusion of vertex corrections may lead to incorrect physical results, especially in relation to insulating gap and pseudogap phenomenology (see Sec.~VII.B of Ref.~\cite{Schaefer-2021} for a recent discussion). 
This is the reason that underlies our choice of the non-self-consistent $t$-matrix approach for the calculation of (frequency-dependent) rf spectra. 

\vspace{-0.3cm}
\subsection{Discovering of a fixed point in the rf spectra}
\label{sec:fixed}

A recent experimental study of the rf spectral intensity of the homogeneous unitary Fermi gas in a box potential has observed a single peak for all temperatures, from the low-temperature superfluid regime to the high-temperature Boltzmann gas regime \cite{Mukherjee-2019}. 
The experiment was performed with a choice of the hyperfine states in $^6$Li which makes negligible the final-state interactions between the state $f$ and the states $(\uparrow,\downarrow)$. 
The presence of a single peak in the spectra at unitarity contrasts with the early observation of two peaks in experiments with inhomogeneous Fermi gases trapped in harmonic potentials (and also in the presence of strong final-state interactions)
\cite{Chin-2004}. 
Our calculations confirm the appearance of a single peak for the homogeneous unitary Fermi gas from the superfluid critical temperature up to the high-temperature Boltzmann regime (see Fig.~\ref{Figure-1}(b)). 
This feature is also found for weaker coupling values, as shown in Fig.~\ref{Figure-1}(a) for $\coup=-0.5$. 
On the other hand, already for the stronger coupling $\coup=+0.5$ two peaks appear in the spectra (as shown in Fig.~\ref{Figure-1}(c)).

\begin{figure}[t]
\begin{center}
\includegraphics[width=8.5cm]{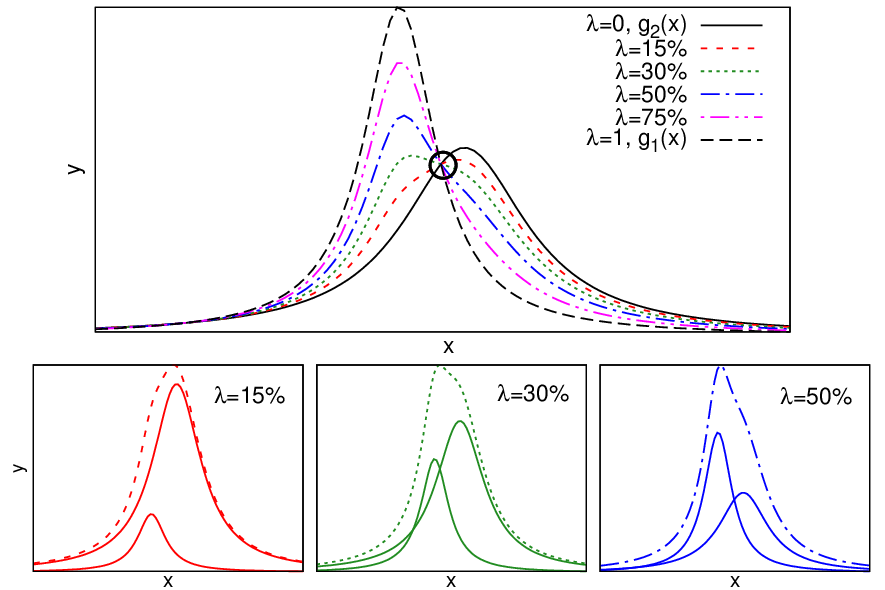}      
\caption{Family of curves $y_\lambda(x)=\lambda g_1(x)+(1-\lambda) g_2(x)$ with varying parameter $0 \le \lambda \le 1$ (top panel). 
              Each curve is analyzed through a linear combination of the two underlying generating curves $g_1(x)$ and $g_2(x)$, as illustrated in the bottom panels for three representative values of $\lambda$.} 
\label{Figure-2}
\end{center}
\end{figure}

An interesting feature, which is common to the temperature evolution of the spectra for all coupling values, is the presence of a ``fixed point" in the spectra. 
Specifically, for given coupling, spectra at different temperatures intersect each other at a point, which is marked by an empty circle in the top panels of Fig.~\ref{Figure-1}.
The presence of this fixed point may be interpreted quite naturally in terms of two peaks that underly \emph{all\/} spectra, even when just a single peak appears in the spectra (which is the case of unitarity and of the BCS side of the crossover).
This interpretation is supported by the generic analysis made in Fig.~\ref{Figure-2}, that shows the family of curves $y_\lambda(x)=\lambda g_1(x)+(1-\lambda) g_2(x)$ with varying parameter $0 \le \lambda \le 1$.
The family is obtained by two generating curves $g_1(x)$ and $g_2(x)$ (in the example of Fig.~\ref{Figure-2}, two Lorentzian curves with different centers and widths are considered). 
One sees that all curves pass through the same fixed point, suggesting that also the rf spectra could be generated by two underlying single-peak structures. 

Out of this proof-of-principle scheme, we have devised a fitting procedure to identify the two components directly from the rf spectra with suitable fitting functions, as described in detail in Sec.~\ref{sec:fitting}.
The outcomes of this procedure for a representative temperature are shown in the bottom panels of Fig.~\ref{Figure-1}. 
Note that, while the two fitting components are clearly visible already in the rf spectra for coupling $\coup=0.5$, at unitarity the two components partially overlap each other producing the shoulder structure visible in the rf spectra. 
On the other hand, at $\coup=-0.5$ no shoulder structure is visible in the rf spectra because one component hides completely the other one.

\vspace{-0.6cm}
\subsection{Fitting procedures of the rf spectra along the BCS-BEC crossover}
\label{sec:fitting}

In this subsection, we outline the procedure that we have implemented for selecting the appropriate fitting functions and the associated fitting procedure to analyze the rf spectra along the BCS-BEC crossover. 

To this end, it is first instructive to examine the retarded pair propagator $\Gamma_0^{\rm R}(\vQ,\Om)$ for real frequency $\Om$.  
Analytic continuation $i\Omega_{\nu}\to\Omega+i0^+$ in the expression (\ref{gamma0}) yields
\begin{align}
&\begin{aligned}
 \Re\left[\Gamma_{0}^{\rm R}(\vQ, \Omega)^{-1}\right] &=\Re\left[t_{2}(\vQ, \Omega)^{-1}\right] \\
& \quad \quad\quad  + {\cal P} \!\! \int \frac{d \vk}{(2 \pi)^{3}} \frac{f\left(\xi_{\vk+\frac{\vQ}{2}}\right)}{\frac{k^2}{2m}-\frac{\Om-\Omth}{2}} 
\label{reinvg0}
\end{aligned}\\
&\begin{aligned}
&\Im\left[\Gamma_{0}^{\rm R}(\vQ, \Omega)^{-1}\right]=\operatorname{Im}\left[t_{2}(\vQ, \Omega)^{-1}\right] \\
& \times \!\! \left[1+\frac{1}{\beta Q \sqrt{\frac{\Omega-\Omega_{\rm th}}{4m}}} \log \left(\frac{1+e^{-\beta\left(\frac{\Om}{2}+Q \sqrt{\frac{\Omega-\Omega_{\rm th}}{4m}}\right)}}{1+e^{-\beta\left(\frac{\Om}{2}-Q \sqrt{\frac{\Omega-\Omega_{\rm th}}{4m}}\right)}}\right)\right] \, ,
\label{iminvg0}
\end{aligned}
\end{align}
where $\cal P$ stands for the principal part value, $\beta = 1/T$ is the inverse temperature, and the off-shell two-body $t$-matrix in vacuum $t_{2}(\vQ, \Omega)$ is given by
\begin{equation}
t_{2}(\mathbf{Q}, \Omega)^{-1}=-\frac{m}{4 \pi a_{\rm F}} + {m^{3/2} \over 4 \pi}\sqrt{\Omth-\Omega-i 0^{+}}  \label{2btm}
\end{equation}
with $\Omth = {Q^2 \over 4m} - 2 \mu$ and $Q = |\mathbf{Q}|$.

\begin{figure}[t]
\begin{center}
\includegraphics[width=9.2cm,angle=0]{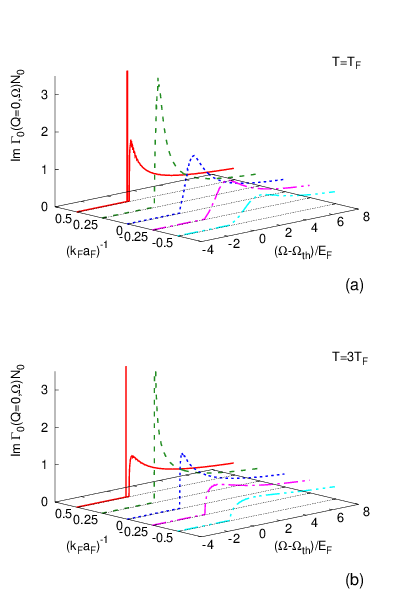}
\caption{Imaginary part of the pair propagator $\Im[\Gamma_0^{\rm R}(\vQ=0,\Om)]$ (in units of $N_0^{-1}$ with $N_0=mk_{\rm F}/(2\pi^2)$) as a function of coupling $\coup$
              and frequency difference $\Om -\Omth$ (in units of $E_{\rm F}$) for (a) $T = T_{\rm F}$ and (b) $T = 3T_{\rm F}$ (where $T_{\rm F}$ is the Fermi temperature).} 
\label{Figure-3}
\end{center}
\end{figure} 

The imaginary part (\ref{iminvg0}) of the retarded pair propagator plays essentially the role of a pair spectral function. 
This can be seen from Fig.~\ref{Figure-3}, which shows $\Im[\Gamma_0^{\rm R}(\vQ,\Om)]$ as a function of coupling $\coup$ and frequency difference $\Om-\Omth$ with center-of-mass momentum $\vQ = 0$ for two representative temperatures 
$T = T_{\rm F}$ and $T = 3T_{\rm F}$. 
When two particles are not bound together, their energy is limited from below by the center-of-mass kinetic energy $\Omth = {Q^2 \over 4m} - 2 \mu$ measured with respect to the single-particle chemical potential.
Bound pairs appear instead below threshold as a Dirac-delta peak, as it is observed in Fig.~\ref{Figure-3} on the BEC side of the crossover $\coup\gtrsim  0.5$.

Moving toward the unitarity regime $0.0\lesssim \coup \lesssim 0.5$, no stable pairs are found unless their kinetic energy or the temperature are sufficiently large, such that
Pauli exclusion effects due to the surrounding medium become irrelevant and  weakly-bound molecules can form as if they were in vacuum.
For smaller momenta or temperatures, $\Re[\Gamma_0^{\rm R}(\vQ,\Om)^{-1}]$ has a zero above the threshold
$\Omth$ of the continuum, leading to a resonance peak of $\Im[\Gamma_0^{\rm R}(\vQ,\Om)]$.
The closer to threshold the resonance is, the more pronounced this peak becomes, up to the point that the zero crosses $\Omth$ and a bound state forms. 
We regard these states close to threshold as quasi-bound states, given their similarity and proximity to stable bound states. 
Physically, they represent unstable and relatively short-lived molecules in the medium.  When crossing unitarity and moving toward the BCS side, they evolve into virtual states \cite{Ma-1953,Nussenzveig-1959,Deltuva-2020} in the medium.
Way above threshold, on the other hand, pair spectral functions share the same high-frequency tail, which represents the spectral distribution of a continuum of two-body scattering states.

The above considerations lead us to identify two different types of two-particle states, namely, bound (B) or quasi-bound (QB) states which occur at low energy (below or close to threshold), and high-energy scattering (S) states.
These different two-particle states have important effects on the fermionic single-particle properties. 
Accordingly, we separate the expression~(\ref{imsancon}) of the single-particle self-energy into a bound (bnd) part originating from the polar contribution
and an unbound (ubn) part originating from the continuum contribution of the pair spectral function $\Im[\Gamma_0^{\rm R}(\vQ,\Om)]$:
\begin{align}
\Im\left[\Sigma^{\rm bnd}(\vk, \om)\right]&=-\int \frac{\mathrm{d} \vQ}{(2 \pi)^{3}}\left[b\left(\omega+\xi_{\vQ-\vk}\right)+f\left(\xi_{\vQ-\vk}\right)\right] \nonumber \\
& \quad\quad \times \Im\left[\Gamma_{0}^{\rm polar}\left(\vQ, \omega+\xi_{\vQ-\vk}\right)\right] \label{sigbnd} \\
\Im\left[\Sigma^{\rm ubn}(\vk, \om)\right]&=-\int \frac{\mathrm{d} \vQ}{(2 \pi)^{3}}\left[b\left(\omega+\xi_{\vQ-\vk}\right)+f\left(\xi_{\vQ-\vk}\right)\right] \nonumber \\
& \quad\quad \times \Im\left[\Gamma_{0}^{\rm cont}\left(\vQ, \omega+\xi_{\vQ-\vk}\right)\right] \label{sigubn}
\end{align}
where by our definition
\begin{align}
&\Im\left[\Gamma_{0}^{\rm polar}(\vQ, \Omega)\right]=\left\{\begin{array}{ll}
\Im\left[\Gamma_{0}^{\rm R}(\vQ, \Omega)\right] & \Omega<\Omega_{\rm th} \\
0 & \Omega>\Omega_{\rm th} \label{img0pole}
\end{array}\right.\\
&\Im\left[\Gamma_{0}^{\rm cont}(\vQ, \Omega)\right]=\left\{\begin{array}{ll}
0 & \Omega<\Omega_{\rm th} \\
\Im\left[\Gamma_{0}^{\rm R}(\vQ, \Omega)\right] & \Omega>\Omega_{\rm th} \, .
\end{array}\right. \label{img0cont}
\end{align}

\begin{figure}[t]
\begin{center}
\includegraphics[width=9.0cm,angle=0]{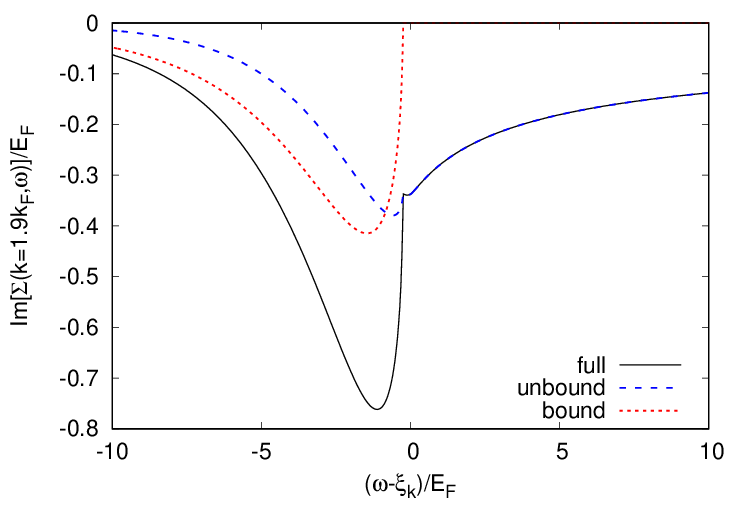}
\caption{Imaginary part of the self-energy at $k/k_{\rm F} = \sqrt{T/T_{\rm F} }\simeq 1.9$ as a function of the frequency difference $\omega- \xik$ (in units of $E_{\rm F}$), for $\coup=0.5$ and $T /T_{\rm F} = 3.5$. 
             The full function Im$\Sigma$ (full line) is separated in its unbound (dashed line) and bound (dotted line) contributions.}  
\label{Figure-4}
\end{center}
\end{figure} 

Figure~\ref{Figure-4} shows  an example of $\Im[\Sigma(\vk, \om)]$ as a function of frequency, for a choice of temperature and coupling such that the discrete and continuum contributions are equally important. 
Here, the wave vector is fixed at a representative thermal value $k/k_{\rm F} = \sqrt{T/T_{\rm F}}$, while the frequency $\omega$ is shifted by the free dispersion $\xik$.

We will now show that one can split the single-particle self-energy into two different contributions corresponding to the above two types of two-particle states. 

\subsubsection{Bound/quasi-bound-state contribution to the self-energy}

In order to disentangle the different contributions to $\Im \Sigma(\vk, \om)$ in the way anticipated above, we first consider the BEC limit in which the pair propagator acquires a simple polar form.
This allows us to derive an analytic expression $\Im \Sigma_{\rm BEC}^{\rm bnd}$ for the BEC limit of the bound contribution to $\Im \Sigma$ (we refer to Appendix~\ref{appA} for details). 
In particular,  two contributions $\Im \Sigma_{\rm BEC}^{\rm bnd} = \Im \Sigma_{\rm B}^{\rm bnd} + \Im \Sigma_{\rm F}^{\rm bnd}$ can be identified, which read:
\begin{align}
\Im \bar{\Sigma}_{\rm B}^{\rm bnd}(\bar{\vk},\bar{\om})&=-\Theta(\bar{\xi}_{\bar{k}}-\bar{\om}-{\bar\varepsilon}_0^*)\frac{W}{16\pi^2\bar{\beta}\bar{k}}\nonumber\\
&\times \log\left(\frac{1-\mathrm{e}^{-\bar{\beta}\left(\bar{\xi}_{\bar{k}}^{+}+\bar{\omega}\right)}}{1-\mathrm{e}^{-\bar{\beta}\left(\bar{\xi}_{\bar{k}}^{-}+\bar{\omega}\right)}}\right) \label{sigbecB} \\
\Im \bar{\Sigma}_{\rm F}^{\rm bnd}(\bar{\vk},\bar{\om})&=-\Theta(\bar{\xi}_{\bar{k}}-\bar{\om}-\bar{\varepsilon}_0^*)\frac{W}{16\pi^2\bar{\beta}\bar{k}}\nonumber\\
&\times \log \!\left(\!\frac{1+\mathrm{e}^{-\bar{\beta}\bar{\xi}_{\bar{k}}^{-}}}{1+\mathrm{e}^{-\bar{\beta}\bar{\xi}_{\bar{k}}^{+}}} \right) . \label{sigbecF} 
\end{align}
Here, the overline signifies that all quantities are dimensionless (to ease comparison with Appendix A), $\Theta$ is the unit step function, $W={64\pi^2 \over k_{\rm F}a_{\rm F}}$ is the pole weight,
$\bar{\xi}_{\bar{k}}^\pm=[\bar{k}\pm\sqrt{2(\bar{\xi}_{\bar{k}}-\bar{\om}-\bar{\varepsilon}_0^*)}]^2-\bar{\mu}$, and the (dimensionless) energy $\bar{\varepsilon}_0^*=\varepsilon_0^*/E_{\rm F}$ reduces to the (dimensionless) two-body binding energy  $\bar{\varepsilon}_0=\be/E_{\rm F}= 2/(k_{\rm F}a_{\rm F})^2$ in the BEC limit.  
Distinguishing $\varepsilon_0^*$ from $\varepsilon_0$ is useful because the expressions (\ref{sigbecB})$-$(\ref{sigbecF}) can be used to describe the bound state contribution to $\Im \Sigma$ even away from the BEC limit, 
provided that $\varepsilon_0^*$ (as well as $W$) are treated as fitting parameters that recover the above values only in the BEC limit.

Considering further the unbound part, one sees from Fig.~\ref{Figure-4} that $\Im \Sigma_{\rm B}^{\rm ubn}$ is peaked at $\om \simeq \xik$ with a different decay on the two sides of the peak, namely, 
exponential for $\om \lesssim \xik$ and algebraic for $\om \gtrsim \xik$.  
One can verify from the expression~(\ref{sigubn}) of the unbound self-energy that single-particle states with frequency $\om \lesssim \xik$ originate mainly from two-particle states of the two-body continuum near threshold, 
while single-particle states with higher frequency $\om \gtrsim \xik$ originate mainly from two-particle scattering states away from threshold. 
Accordingly, we may associate the range $\om \lesssim \xik$ with quasi-bound (QB) states and the range $\om \gtrsim \xik$ with scattering (S) states. 

Note also from Fig.~\ref{Figure-4} that the range $\om \lesssim \xik$ associated with QB states almost coincides with the range where the bound state contribution associated with B states is present, 
and that for $\om\ll\xik$ the unbound part shares the same asymptotic behavior of the bound part.
Taking advantage of this similarity, the analytic expressions~(\ref{sigbecB}) and (\ref{sigbecF}) can be considered appropriate fitting functions to describe both B and QB states.  
After applying the fitting procedure, we will actually verify that B states are well described by the contribution~(\ref{sigbecB}) and QB states by the contribution~(\ref{sigbecF}).

\subsubsection{Scattering states contribution to the self-energy}

We are now left with the last energy sector $\om \gtrsim \xik$ associated with the high-energy S states. 
In this case, we focus on the continuum part of the pair propagator~(\ref{img0cont}), corresponding to the range of frequencies $\Om>\Omth$. 

We first note that in this frequency range $\Re[t_{2}(\vQ, \Omega)^{-1}]=-{m k_{\rm F} \over 4\pi} \coup$.
For temperatures not too close to $T_c$, one can neglect the wave vector and frequency dependence of $\Re[\Gamma_{0}^{R}(\vQ,\Om)^{-1}]$, effectively reducing it to a shift of the coupling constant appearing in $\Re[t_{2}(\vQ, \Omega)^{-1}]$.
Accordingly, we can approximate  $\Re[\Gamma_{0}^{\rm R}(\vQ,\Om)^{-1}]\simeq -{m k_{\rm F} \over 4\pi} (k_{\rm F}a_{\rm F})^{-1}_{\rm eff}$, while retaining the full wave vector and frequency dependence of 
$\Im[\Gamma_{0}^{\rm R}(\vQ,\Om)^{-1}]$ given by Eq.~(\ref{iminvg0}). 
For clarity, we rename the propagator obtained in this way $\Gamma_{0}^s(\vQ,\Om)$.
The effective coupling constant $(k_{\rm F}a_{\rm F})^{-1}_{\rm eff}$ then constitutes a further parameter of the fitting procedure.

Once the above pair propagator is inserted in the expression~(\ref{sigubn}) of self-energy, it produces a continuum of unbound states that includes both quasi-bound and scattering states. 
Quasi-bound states, on the other hand, have been already taken into account by the fitting function~(\ref{sigbecF}).
We thus multiply the right-hand side of Eq.~(\ref{sigubn}) of the unbound self-energy by the Fermi function
\begin{equation}
F(\mathbf{k}, \omega)=\frac{1}{\exp \left[\left(\xi_{\mathbf{k}}-\varepsilon_0^*-\omega_{s}(k)-\omega\right) / T_{s}\right]+1} \label{fermilike} \, ,
\end{equation}
which selects states with energy above the boundary $\xik-\omth$ between QB and S states. 
The boundary value is then adjusted by adding a wave vector dependence $\om_s(k) = \alpha\,k^2$ so as to make the merging of the two (QB and S) sectors as smooth as possible, where $\alpha$ and $T_s$ are additional fitting parameters.

\subsubsection{Overall fitting procedure for the rf spectra}

\begin{figure*}[t]
\begin{center}
\includegraphics[width=\textwidth]{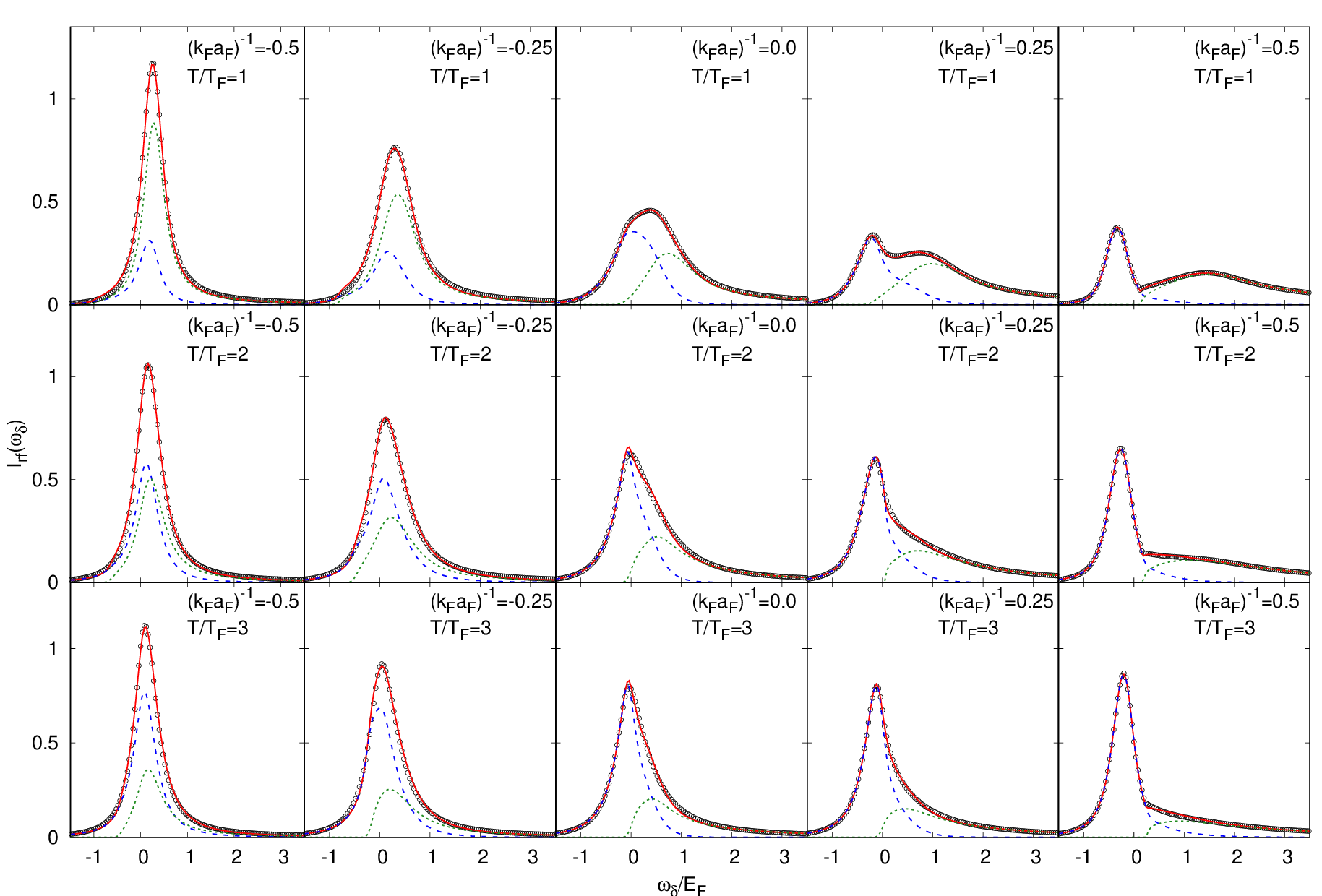}
\caption{Temperature and coupling evolution of the two fitting components to the dimensionless rf spectral intensity $I_{\rm rf}$ throughout the BCS-BEC crossover.
              The sum of the bound/quasi-bound (dotted line) and scattering (dashed line) contributions gives rise to the function (full line) that fits the calculated rf  spectral intensities $I_{\rm rf}$ (circles) 
              as a function of the detuning frequency $\omega_{\delta}$ (in units of $E_{\rm F}$).} 
\label{Figure-5}
\end{center}
\end{figure*} 

By putting together all the above approximate expressions for the different contributions to $\Im\Sigma(\vk,\omega)$, we obtain eventually the semi-analytic function
$S(\vk,\om)$ 
\begin{align}
S(\vk, \omega) &=S_{\rm B/QB}(\vk, \omega)+S_{\rm S}(\vk, \omega) \label{fitfunc} \\
S_{\rm B/QB}(\vk, \omega) &=\Im \Sigma_{\rm B}^{\rm bnd}(\vk, \omega)+\Im \Sigma_{\rm F}^{\rm bnd}(\vk, \omega)  \label{fitbnd} \\
S_{\rm S}(\vk, \omega) &=U(\vk, \omega) F(\vk, \omega) \label{fitscatt}
\end{align}
with $\Im \Sigma_{\rm B}^{\rm bnd}(\vk, \omega)$ and $\Im \Sigma_{\rm F}^{\rm bnd}(\vk, \omega)$ given by Eqs.~(\ref{sigbecB}) and (\ref{sigbecF}), respectively, and 
\begin{align}
U(\vk, \omega)=-\int \frac{\mathrm{d} \vQ}{(2 \pi)^{3}} & \Im\left[\Gamma_{0}^{s}\left(\vQ, \omega+\xi_{\vQ-\vk}\right)\right] \times \\
& \times\left[b\left(\omega+\xi_{\vQ-\vk}\right)+f\left(\xi_{\vQ-\vk}\right)\right],
\end{align}
with $F(\vk,\om)$ given by Eq.~(\ref{fermilike}).
The component $S_{\rm B/QB}$ is meant to represent both bound B and quasi-bound QB states (described by $\Im\Sigma_{\rm B}^{\rm bnd}$ and $\Im \Sigma_{\rm F}^{\rm bnd}$, respectively) with $\om \lesssim \xik-\omth$. 
The scattering component $S_{\rm S}$ is meant to represent a continuum of scattering states with $\om\gtrsim\xik-\omth$. 

The function $S(\vk,\om)$, that depends on the five parameters $(W, \omth, (k_{\rm F}a_{\rm F})^{-1}_{\rm eff},\alpha,T_s)$, is then used to fit the calculated rf spectra as follows:
\begin{enumerate}[(i)]
\item From the approximate expression $S(\vk,\om)$ for $\Im\Sigma(\vk,\om)$, a corresponding  approximate expression for $\Re\Sigma(\vk,\om)$ is obtained via a Kramers-Kronig transform and the single-particle spectral function 
$A(\mathbf{k},\omega)$ given by Eq.~(\ref{akwdef}) is calculated;
\item The latter is then plugged into the expression (\ref{irfnorm}) to obtain the rf spectral intensity $I_{\rm fit}(\om)$;
\item A least-square method is used to obtain the best values of the five fitting parameters, where  $\chi^2 =\sum_{i} [I_{\rm rf}(\om_i)-I_{\rm fit}(\om_i)]^2$ is the quantity to be minimized.
Here, the frequencies $\{\omega_i, i=(1,\cdots,N)\}$ are chosen on a sufficiently dense grid, so to follow all the features of the curve $I(\om_i)$.  (Typically, we take $N=1000$ in the interval $[-20 E_{\rm F}, 20 E_{\rm F}]$.)
\end{enumerate}

Recall in this context that the thermodynamic parameters $(\mu,T)$ entering the expression of $S(\vk,\om)$ are not fitting parameters. 
The temperature $T$ is set as an input parameter at the outset of the numerical calculation, while the chemical potential $\mu$ is obtained by solving the number equation (\ref{densequ}).

The outcome of the above fitting procedure is shown in Fig.~\ref{Figure-5} for a number of representative values of coupling and temperature.  
One sees that the fitting procedure reproduces quite well the numerical rf curves in all cases here considered, revealing at the same time the underlying two-peak structure of the spectra as we had anticipated in subsection~\ref{sec:fixed}. 
Depending on the values of temperature and coupling, one peak may become dominant over the other one.
For fixed coupling, the scattering component~(\ref{fitscatt}) becomes progressively more important as the temperature increases, eventually reaching the unitary Boltzmann-gas regime $T\gg\be$ (see Appendix~\ref{appD1}),
which is characterized in practice by a single peak.
For fixed temperature, by increasing the coupling toward the BEC side ($\coup \gg T/T_F$) the bound component dominates, as expected for a gas of bosonic molecules (see Appendix~\ref{appB}). 
A single peak dominates also in the Fermi liquid regime $|\coup| \gg T/T_F$ with $\coup<0$ (not shown in Fig.~\ref{Figure-5}).

It is, finally, worth mentioning that in appropriate regimes our fitting procedure can be related to the fitting scheme introduced in Ref.~\cite{Sagi-2015} to interpret momentum-resolved rf spectroscopy. 
Specifically, in Appendix~\ref{appB} we show how the bound part of our fitting spectral function reduces to the incoherent part of the spectral function of Ref.~\cite{Sagi-2015}, both in the BEC and high-temperature regimes.

\begin{figure}[t]
\begin{center}
\includegraphics[width=9.cm]{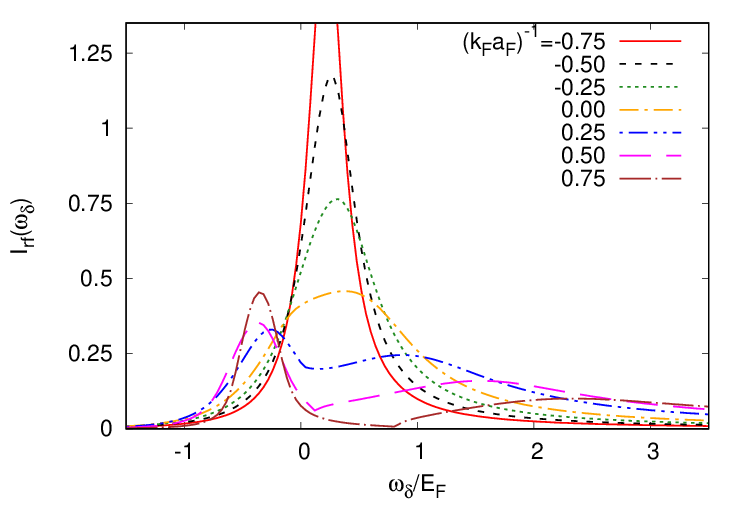}
\caption{Dimensionless rf spectral intensity $I_{\rm rf}$ as a function of detuning frequency $\omega_{\delta}$ (in units of $E_{\rm F}$) at $T = T_{\rm F}$ for several values of the coupling $\coup$. }
\label{Figure-6}
\end{center}
\end{figure} 

\subsection{Widths of rf peaks and pair size}
\label{sec:numres}

The evolution of the rf spectra with coupling at fixed temperature shown in Fig.~\ref{Figure-6} emphasizes  the gradual formation of the double-peak structure discussed above. 
Specifically, while for negative couplings a single peak is present, for  positive couplings two peaks are clearly visible. 
For the latter couplings, the peak at negative detuning frequency $\omega_{\delta}$ (roughly) corresponds to a nearly-free fermion added to the system, while the peak at positive detuning frequency is associated with fermions excited from pairs.
 
To deepen the analysis of the rf spectra, we have determined the full-width at half-maximum (FWHM) $E_{\rm w}$ of the peaks in the spectra. 
In particular, for couplings from weak to unitarity we have calculated $E_{\rm w}$ for the single peak which is present in the rf spectra, while for positive couplings we have separately considered $E_{\rm w}$  for both peaks. 
In this second case, the fitting procedure discussed above has been crucial to disentangle the contribution of the two peaks.

\begin{figure}[t]
\begin{center}
\includegraphics[width=9.0cm,angle=0]{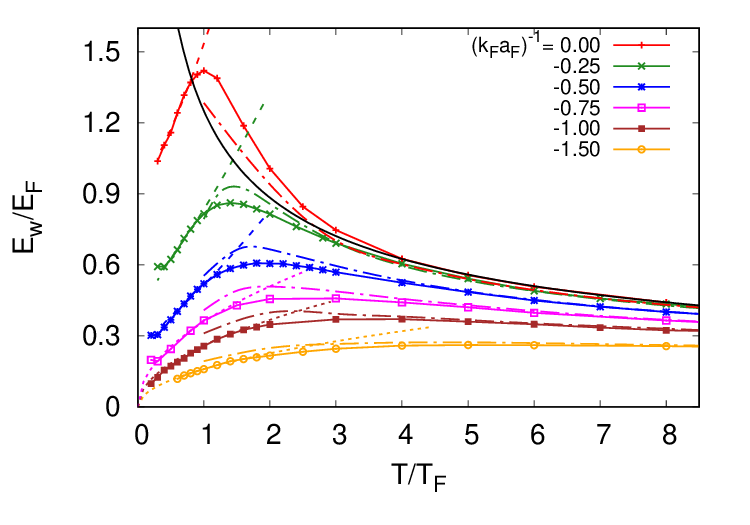}
\caption{FWHM $E_{\rm w}$ (in units of $E_{\rm F}$) as a function of temperature on the negative coupling side of the crossover (full lines with symbols). 
              Dashed lines: fits to the low-temperature dependence of $E_{\rm w}$ with the expression $B/ (2m \xi_{\rm pair})^2$, where $B$ is a constant and $\xi_{\rm pair}$ is the calculated temperature-dependent pair size (see below). 
              Dotted lines: fits to the low-temperature dependence of  $E_{\rm w}/E_{\rm F}$ with  $B' \sqrt{T/T_{\rm F}}$,  where $B'$ is a constant. 
              Dashed-dotted lines: high-temperature benchmarks for $E_{\rm w}$ obtained in the Boltzmann limit when $e^{\beta \mu} \ll 1$. 
              Full black line: extremely asymptotic high-temperature result  $E_{\rm w}/E_{\rm F}=1.25 \, (T/T_{\rm F})^{-1/2}$.}
\label{Figure-7}
\end{center}
\end{figure} 

\begin{figure}[h]
\begin{center}
\includegraphics[width=8.cm,angle=0]{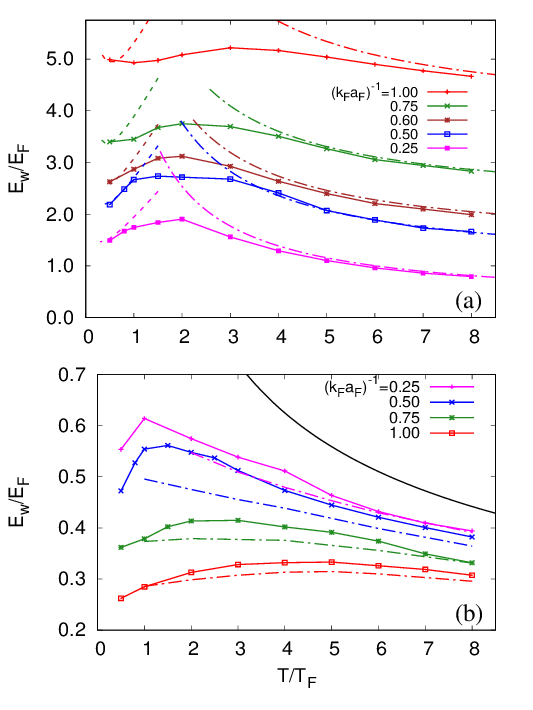}
\caption{FWHM $E_{\rm w}$ (in units of $E_{\rm F}$) as a function of temperature on the positive coupling side of the crossover.
             (a): $E_{\rm w}$ for the B/QB (pairing) peak (full lines with symbols); fits to the low-temperature dependence of $E_{\rm w}$ with the expression $B/ (2m \xi_{\rm pair})^2$ (dashed lines); high-temperature benchmarks (dashed-dotted lines).
             (b): $E_{\rm w}$ for the S (quasi-particle) peak (full lines with symbols); high-temperature benchmarks (dashed-dotted lines); 
             extremely asymptotic high-temperature result $E_{\rm w}/E_{\rm F}=1.25 \, (T/T_{\rm F})^{-1/2}$ (full black line).}
\label{Figure-8}
\end{center}
\end{figure} 

The resulting widths $E_{\rm w}$ are shown in Figs.~\ref{Figure-7} and~\ref{Figure-8} as a function of temperature for a number of representative couplings in the crossover region. 
In particular, Fig.~\ref{Figure-7} corresponds to negative couplings for which a single peak is present, while Fig.~\ref{Figure-8} corresponds to positive couplings for which two peaks are present.
Accordingly, Fig.~\ref{Figure-8} is split in two panels that separately report $E_{\rm w}$ for (a) the B/QB peak (which can be interpreted as a ``pairing peak" in this coupling range) and (b) the S peak (which can be interpreted as a ``quasi-particle peak" in this coupling range).

In this context, we recall that Ref.~\cite{Schunck-2008} has proposed to correlate the width $E_{\rm w}$ of the rf spectra to the Cooper pair size $\xi_{\rm pair}$ in the superfluid phase. 
Specifically, by calculating rf spectra and thermodynamic quantities within the BCS theory at zero temperature, it was found in Ref.~\cite{Schunck-2008} that the width $E_{\rm w}$ is proportional to the energy scale $1/(2m \xi_{\rm pair}^2)$, 
with $\xi_{\rm pair}$ calculated like in Refs.~\cite{Pistolesi-1994,Marini-1998}. 
In addition, the constant of proportionality $B$ between $E_{\rm w}$ and $1/ (2m \xi_{\rm pair}^2)$ was found to be relatively weakly dependent on coupling (increasingly monotonically from about 0.3 in the BCS limit to about 1.9 in the BEC limit~\cite{Schunck-2008}). 
The experimental spectra reported in the same work at low temperature in the superfluid phase have confirmed qualitatively this kind of behavior, by finding a progressively increasing $E_{\rm w}$ from the BCS to the BEC regimes
which reflects the expected shrinking of the pair size \cite{Pistolesi-1994,Marini-1998}.

\begin{figure}[t]
\begin{center}
\includegraphics[width=8.cm,angle=0]{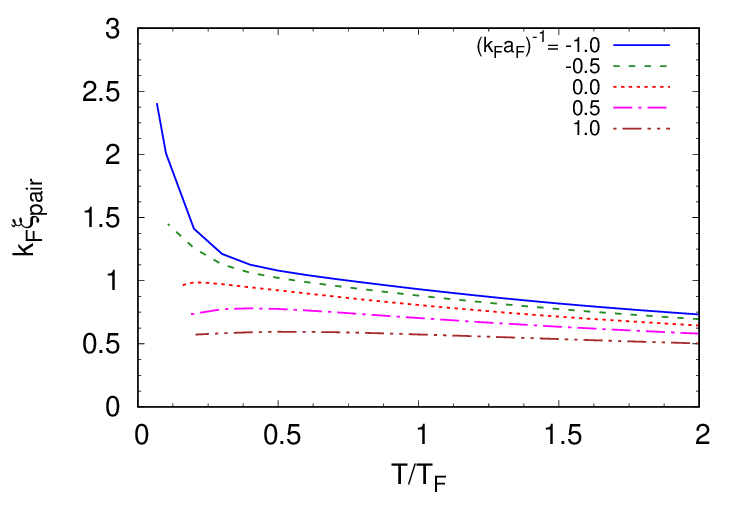}
\caption{Pair size (or pair correlation length) $\xi_{\rm pair}$ (in units of $k_{\rm F}^{-1}$) as a function of temperature (in units of $T_{F}$) for different couplings throughout the BCS-BEC crossover.}
\label{Figure-9}
\end{center}
\end{figure} 

We are here in a position to extend the comparison between the width $E_{\rm w}$ of the rf spectra and the energy scale $1/(2m \xi_{\rm pair}^2)$ to {\em finite temperature in the normal phase\/}.  
Such an extension requires an appropriate definition of the pair size $\xi_{\rm pair}$ in the normal phase, for which one cannot rely on a BCS formalism like in Ref.~\cite{Schunck-2008}. 
To this end, we rely on the definition of the pair size given in Ref.~\cite{Pistolesi-1996} 
\begin{equation}
\xi_{\rm pair}^2 = \frac{\int \! d {\bf r}  g_{\uparrow\downarrow}({\bf r}) {\bf r}^2}{\int \! d {\bf r}  g_{\uparrow\downarrow}({\bf r})} \, ,
\label{xi-from-g}
\end{equation} 
which is based on the knowledge of the pair correlation function 
\begin{equation}
g_{\uparrow\downarrow}({\bf r})=\langle {\hat \psi}_{\uparrow}^{\dagger}({\bf r}){\hat \psi}^{\dagger}_{\downarrow}(0) {\hat \psi}_{\downarrow}(0) {\hat \psi}_{\uparrow}({\bf r})\rangle -(n/2)^2 
\end{equation}
that can be used to calculate $\xi_{\rm pair}$ also in the normal phase~\cite{Palestini-2014}. 
As anticipated in subsection~\ref{sec:nsctheory}, we find it appropriate to calculate a non-dynamical quantity like the pair correlation function $g_{\uparrow\downarrow}({\bf r})$ in terms of the fully self-consistent $t$-matrix approach as detailed in Ref.~\cite{Pini-2020} at the single-particle level, which is consistent with the prescriptions by Kadanoff and Baym\cite{Baym-1961,Baym-1962}. 

The corresponding results for $\xi_{\rm pair}$ are shown in Fig.~\ref{Figure-9} as a function of temperature for several couplings.
These results for $\xi_{\rm pair}$ can then be used to obtain the energy scale $1/(2m \xi_{\rm pair}^2)$, to be compared with the width $E_{\rm w}$ of the overall single peak of the rf spectra from weak coupling to unitarity
as shown in Fig.~\ref{Figure-7}, or of the B/QB (pairing) peak for positive couplings past unitarity as shown in Fig.~\ref{Figure-8}(a).  
One sees from these figures that in the unitarity regime $-0.5 \lesssim \coup \lesssim 0.5 $ the quantity $B/(2m \xi_{\rm pair}^2)$ follows closely the width $E_{\rm w}$ even in the normal phase here considered, up to temperatures 
as high as $T_{\rm F}$ (which is significantly higher than the critical temperature $T_c$).
The fitting parameter $B$ is found to increase monotonically from 0.4 at $\coup= -0.5$ to 1.3 at $\coup = + 0.5$, in line with the mean-field results at zero temperature presented in Ref.~\cite{Schunck-2008}.   

Just about the upper limit of the unitarity regime, the low-temperature dependence of $E_{\rm w}$ is seen from Fig.~\ref{Figure-8}(a) to change its concavity, from negative for $\coup<0.6$ to positive for $\coup>0.6$. 
Interestingly, the value $\coup=0.6$ corresponds to the coupling strength that separates the pseudo-gap from the molecular phase, as identified in Ref.~\cite{Perali-2011} by the disappearance of an underlying Fermi surface (see in particular Ref.~\cite{Palestini-2014} for a precise identification of such a coupling strength).
Above this value of the coupling strength, $E_{\rm w}$ is seen from Fig.~\ref{Figure-8}(a) to remain essentially flat over a wide temperature range, which becomes progressively more extended as the coupling increases.  
This finding is consistent with one's expectation that in the molecular limit the width $E_{\rm w}$ of the pairing peak should be determined by the binding energy $\epsilon_0$ over an extended temperature range.  
The energy scale $1/(2m \xi_{\rm pair}^2)$ has instead a more pronounced temperature dependence even in the molecular limit, thereby disrupting the correspondence with $E_{\rm w}$ in this limit.
On physical ground, this is due to the fact $\xi_{\rm pair}$ extracted from $g_{\uparrow\downarrow}({\bf r})$ like in Eq.~(\ref{xi-from-g}) includes also some degree of bosonic correlations between different pairs, which become increasingly important in the molecular limit. 
The width $E_{\rm w}$ is instead less sensitive to bosonic correlations, being determined by a fermionic quantity such as the single-particle spectral function. 

Also in the weak-coupling regime $\coup < -0.5$ of the normal phase, the identification of $E_{\rm w}$ with $B/(2m \xi_{\rm pair}^2)$ appears no longer possible. 
This is actually not surprising, since in the BCS regime pairing correlations are expected to be established only below $T_c$.  
In this weak-coupling regime, we have instead found that, from $T_c$ up to temperatures of the order of $1/(m a_{\rm F}^2)$, the width $E_{\rm w}$ can be fitted by the expression $E_{\rm w}/E_{\rm F}=B'\sqrt{T/T_{\rm F}}$ 
(cf. dotted lines in Fig.~\ref{Figure-7}). 
This square root behavior of $E_{\rm w}(T)$, which can be proven analytically when the Boltzmann limit $z=e^{\beta \mu} \ll 1$ and weak-coupling condition $T \ll  1/(m a_{\rm F}^2)$ (with $a_{\rm F} < 0$) are simultaneously satisfied
(see Appendix~\ref{appD2}), is found in Fig.~\ref{Figure-7} to hold even down to $T_c$ deep in the quantum regime. 

The dashed-dotted lines in Figs.~\ref{Figure-7} and \ref{Figure-8} provide a high-temperature benchmark to our numerical calculations. 
Specifically, in Figs.~\ref{Figure-7} and \ref{Figure-8}(b) the benchmark is obtained by calculating the rf signal using the lowest-order expansion of the $t$-matrix self-energy in powers of the fugacity $z$, 
namely, by using expression (\ref{img0vac}) for Im$\Gamma_0^{\rm R}$ and replacing the Fermi and Bose functions appearing in Eq.~(\ref{imsancon}) by Boltzmann functions (we refer to Ref.~\cite{Leyronas-2015} 
for a systematic discussion of the high-temperature expansion of the self-energy). 
For the high-temperature benchmark in Fig.~\ref{Figure-8}(a), on the other hand, an expansion that focused on this minor feature is required since the pairing peak is a minor and sub-leading feature of the rf signal at high temperature. 
This expansion is performed in Appendix~\ref{appC2}, leading to the expression [cf.~Eq.~(\ref{asymBEC})]: 
\begin{equation}
E_{\rm w}=1.89 \left(\be + C \frac{E_{\rm F}^2}{T}\right).
\label{empirical}
\end{equation}
In Appendix~\ref{appC2} the result $C=32/(9\pi k_{\rm F} a_{\rm F})$ for the coefficient $C$ is obtained in the limit $T\gg \be \gg 1$.  
We have empirically found that the high-temperature expression (\ref{empirical}) holds even for small values of $\be$, provided the coefficient $C$ is interpreted as a fitting parameter. 
The dashed-dotted lines in Fig.~\ref{Figure-8}(a) obtained in this way is indeed in good agreement with our numerical results for $E_{\rm w}$. 
Note that the high-temperature limit  $E_{\rm w}= 1.89 \be$ of the expression (\ref{empirical}) corresponds exactly \cite{Schirotzek-2008} to the FWHM obtained from the rf line shape of Feshbach molecules in vacuum \cite{Chin-2005}, 
which also coincides with the BEC limit of the rf signal obtained from BCS theory at zero temperature \cite{Perali-2008,Schirotzek-2008}. 

The point that we would like to emphasize here is that our two-peak analysis of the rf spectra allows us to isolate at all temperatures the feature produced by bound molecules, even when their number, and thus their contribution to the rf spectra, becomes progressively small due to their thermal dissociation. 
This small fraction of molecules will be essentially non-interacting with the medium due to the large kinetic energy associated with their center-of-mass motion at high temperature. 
It is for this reason that their contribution to the rf signal recovers the line shape of Feshbach molecules in vacuum. 

Finally, the full black line in Figs.~\ref{Figure-7} and \ref{Figure-8}(b) corresponds to the universal result $E_{\rm w}/E_{\rm F}=1.25 (T_{\rm F}/T)^{1/2}$, which is expected to be reached by the quasi-particle peak in the Boltzmann limit with the further condition  $T \gg 1/(m a_{\rm F}^2)$ (see Appendix \ref{appD1}). 
One sees from Figs.~\ref{Figure-7} and \ref{Figure-8} that, apart from unitarity and the nearby coupling $\coup=-0.25$, this result provides an extreme asymptotic approximation which is reached at very high temperatures only. 
At unitarity, on the other hand, this result accounts for the temperature dependence of $E_{\rm w}$ already at temperatures a few times $T_{\rm F}$. 
This is in agreement with the experimental finding of Ref.~\cite{Mukherjee-2019}, where the high-temperature expression $E_{\rm w}/E_{\rm F}=1.2 (T_{\rm F}/T)^{1/2}$ was found to compare well with experimental data already at 
$T\simeq 2 T_{\rm F}$. 

The above high-temperature result can be understood in terms of the kinetic theory of gases in the following way.
Within a relaxation time approximation \cite{Smith-1989}, one obtains for the relaxation time $\tau=1/(n \sigma \langle v \rangle)$, where $ \langle v \rangle \propto \sqrt{T/m}$ is the thermally averaged velocity of the atoms
and $\sigma \propto  [{1/a_{\rm F}^{2}}+k_T^2]^{-1} $ is an effective cross section.
When $T\gg T_{\rm F}$, the wave vector $k_T$ can be approximated by its thermal expression $k_T \propto \sqrt{m T}$.
In addition, the width $E_{\rm w}/E_{\rm F}\propto \tau^{-1}$ of the rf spectra is  proportional to the inverse relaxation time. 
In this way, one obtains:
\beq
E_{\rm w}/E_{\rm F} \propto {\sqrt{T/T_{\rm F}} \over (k_{\rm F}a_{\rm F})^{-2}+T/T_{\rm F} } .
\label{kinetic}
\eeq
It is interesting to note that the expression (\ref{kinetic}) recovers both the (extreme) high-temperature behavior $E_{\rm w} \propto 1/\sqrt{T}$ as well as the low-temperature behavior $E_{\rm w} \propto \sqrt{T}$, and that it locates the position of the maximum of $E_{\rm w}(T)$ at $T/T_{\rm F} \propto (k_{\rm F}a_{\rm F})^{-2}$ consistently with what is reported in Figs.~\ref{Figure-7} and \ref{Figure-8}(b).

As a final remark, we mention that under the typical conditions of the rf-spectroscopy experiments, the separation between different hyperfine levels for, e.g., $^6$Li atoms is about $80$ MHz while typical Fermi energies are of the order of 10 kHz \cite{Ketterle-2008,Mukherjee-2019}. This implies that, even at the highest temperatures (of the order of $10 T_{\rm F}$) considered in the present work, thermally induced transitions between different hyperfine levels are completely frozen.

\subsection{Weights of the rf peaks}
\label{sec:weights}

\begin{figure}[t]
\begin{center}
\includegraphics[width=9.cm,angle=0]{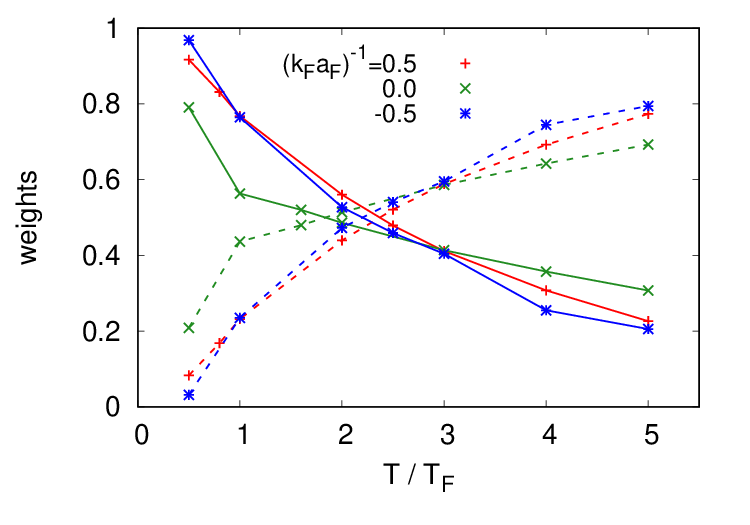}
\caption{Weights of the B/QB and S contributions to the rf spectra (full line and dashed line with symbols, respectively) as a function of temperature for three different couplings.} 
\label{Figure-10}
\end{center}
\end{figure} 

The weights of the two peaks extracted from the rf spectra can also be determined by using our fitting procedure. 
These weights are obtained by calculating the integral over the frequencies, separately for each of the two fitting functions. 

Figure~\ref{Figure-10} shows the results of this calculation for three couplings about unitarity.
As a general trend, we observe that the B/QB contribution, which is dominant at low temperature, progressively gives the way to the S contribution at larger temperature. 
The temperature at which the two contributions exchange their relative importance is smallest at unitarity (with a value slightly below $2 T_{\rm F}$) and progressively increases on both sides of unitarity. 

It might at first appear surprising that the temperature range, within which the B/QB contribution is dominant, somewhat increases its extension with respect to unitarity not only when moving into the BEC regime but also in the BCS regime. 
However, this finding can be explained as follows.
Recalling that the B/QB contribution is associated with the low-energy part of the pair propagator, in the BEC regime this corresponds to the isolated pole associated with molecular binding, 
such that the B/QB contribution can be interpreted as a pairing peak. 
In the BCS regime where this pole is not present, on the other hand, the low-energy part of the pair propagator corresponds to the region which is complementary to the high-energy/high-frequency tail of the pair propagator. 
Owing to the presence of the term $m/(4\pi a_{\rm F})$ in the expression (\ref{gamma0}) of the pair propagator, this tail is pushed to increasingly high-frequencies as the scattering length $a_{\rm F} \to 0$ in the weak-coupling limit, 
thereby enlarging the region in which the B/QB contribution is dominant.

\section{rf spectra and phase diagram} 
\label{sec:phdiag}

\begin{figure}[t]
\begin{center}
\includegraphics[width=8.6cm,angle=0]{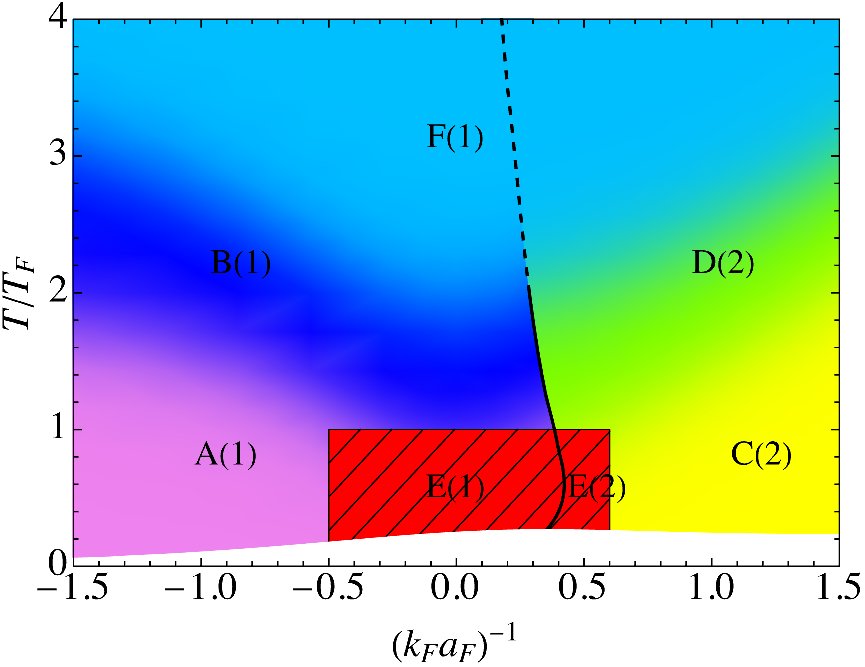}
\caption{Temperature-vs-coupling phase diagram encompassing the BCS-BEC crossover of an ultra-cold Fermi gas, organized into sectors (from A to F) whereby the rf spectra acquire distinctive features that reveal their underlying 
             dominant physical origin. 
             Numbers within parentheses indicate the number of visible peaks in the rf spectra for the corresponding sector.
             Black line (full when it separates two sectors/sub-sectors, dashed otherwise): a dispersive pole appears in the retarded pair propagator $\Gamma^{\rm R}$.
             Blank area at low temperature: superfluid phase, not considered in the present work.} 
\label{Figure-11}
\end{center}
\end{figure} 

The results that we have obtained by our fitting procedure, for the widths and weights of the two peaks emerging from the rf spectra, allow us to identify different physical sectors in the temperature-vs-coupling phase diagram of an
ultra-cold Fermi gas. 
The final outcomes of this analysis are summarized in Fig.~\ref{Figure-11}. 
We now describe in detail this kind of partitioning of the phase diagram, by illustrating the physical features associated with the different sectors of Fig.~\ref{Figure-11}.

In general, the evolution from one to another sector turns out to be a smooth crossover, and as such it is represented in Fig.~\ref{Figure-11} by a continuous evolution from one to another color \cite{footnote-phasediagram}. 
Note also that the number appearing within parentheses in the label of a given sector indicates whether one or two peaks are {\em visible\/} in the ``bare'' rf spectra (that is, before our two-peak fitting procedure is implemented on these spectra). 

\subsection{Sector A: weakly-interacting Fermi gas} 
\label{sectA}

In this sector, the rf spectra show a \emph{single\/} peak with Lorentzian shape, that can be associated with the presence of well-defined quasi-particles. 
In particular, these quasi-particles correspond to a weakly-interacting Fermi gas, which is characterized by the width $E_{\rm w}$ of the peak increasing with temperature as $\sqrt{T/T_{\rm F}}$. 
As discussed in Sec.~\ref{sec:theoretical_approach} (cf. in particular the dotted lines in Fig.~\ref{Figure-7}), this behavior occurs on the weak-coupling side of the BCS-BEC crossover (that is, for $\coup \lesssim -0.5$) and from low to intermediate temperatures (with respect to $T_{F}$). 
In particular, the upper temperature boundary of this sector, which is set by $E_{\rm w}(T)$ departing from its low-temperature benchmark $\propto \sqrt{T/T_{\rm F}}$, increases quite rapidly when going deeper in the weak-coupling regime.  
The analytic calculation reported in Appendix~\ref{appD2} in the weak-coupling limit shows that such a square-root behavior holds for $T_{\rm F} \ll T \ll (m a_{\rm F})^{-2}$, thus explaining the rapid increase of the upper boundary 
of this sector when progressing toward the weak-coupling limit.

\subsection{Sector B: weakly-interacting Boltzmann gas}
\label{sectB}

In this sector, the rf spectra show a \emph{single\/} peak, whose width $E_{\rm w}$ develops a maximum as a function of temperature that becomes increasingly broader and less pronounced as the coupling decreases. 
This sector corresponds to a crossover region between the low- and high-temperature benchmarks of $E_{\rm w}(T)$ discussed in Sec.~\ref{sec:numres}. 
Given its temperature and coupling ranges, this sector can be identified as a weakly-interacting Boltzmann gas. 

Specifically, the lower boundary of sector B for coupling $\coup \le -0.5$ is determined by the upper boundary of the sector A just discussed, while for coupling $\coup > -0.5$ the boundary with sector E is set by the condition $T = T_{\rm F}$ 
that separates the Boltzmann from the quantum degenerate regime. 
The right boundary of sector B is determined by the appearance of a visible second peak in the rf signal (which marks the access to sector D). 
In practice, the (sharp) boundary with sector D as identified by the black line in Fig.~\ref{Figure-11} is determined by the appearance of a dispersive pole in the retarded pair propagator $\Gamma^{\rm R}$, which reflects the formation of molecules in a many-body environment. 
The region to the right of this black line is where such a molecular state is found for all values of the pair momentum ${\bf Q}$ (see Appendix~\ref{appA} for details).
Finally, the upper boundary of sector B is determined by the condition that the B/QB and S contributions to the rf spectra have equal weights.

\subsection{Sector C: weakly-interacting Bose gas} 
\label{sectC}

In this sector, the rf spectra show \emph{two\/} peaks, which are approximately separated by the two-fermion binding energy $\varepsilon_{0}$. 
In addition, the peak at larger detuning frequencies is dominant and can be interpreted as a pairing (molecular) peak.
This sector thus corresponds to the region of the phase-diagram that can be effectively associated with a weakly-interacting Bose gas.
In this sector, the width $E_{\rm w}$ of the pairing peak is essentially flat as a function of temperature and roughly proportional to the binding energy $\varepsilon_{0}$, as discussed in Sec.~(\ref{sec:numres}).

Specifically, the upper boundary of sector C is set by determining when the weight of the bosonic contribution to the pairing peak equals the rest of the spectral weight of $I_{\rm rf}$, 
thereby indicating the presence of a significant number of unbound fermions above this boundary.
The boundary with sector E is given by the vertical line at $\coup=0.6$, which (as mentioned above) corresponds to the coupling strength above which the width $E_{\rm w}$ becomes essentially flat as a function 
of temperature and proportional to $\varepsilon_{0}$.
 
\subsection{Sector D: fermion-dimer mixture}
\label{sectD}

In this sector, the rf spectra show \emph{two\/} peaks approximately separated by the two-fermion binding energy $\varepsilon_{0}$, similarly to sector C. 
However, in sector D the pairing (molecular) peak is no longer dominant over the quasi-particle peak associated with unbound fermions.
Sector D thus corresponds to a fermion-dimer mixture, where bosonic dimers coexist with their fermionic constituents. 
This mixture provides an intermediate state between a gas of bosonic molecules (dimers) occurring at low temperature, and a completely dissociated classical gas of the constituent atoms occurring at high temperature.

Like for sector B, the upper boundary of sector D is determined by the criterion that 
the B/QB and S contributions to the rf spectra have equal weights,
signaling entrance to the classical regime.
 
\subsection{Sector E: degenerate unitary Fermi gas region} 
\label{sectE}

In this sector, information on the pair size $\xi_{\rm pair}$ can be extracted from the rf spectra, thereby extending to the normal phase what was previously found to hold in the superfluid phase at $T=0$ \cite{Schunck-2008}.
Specifically, as discussed in Sec.~\ref{sec:numres}, the single peak which is visible in the sub-sector E(1), or the pairing peak in the sub-sector E(2) where two peaks are instead visible, displays a width $E_{\rm w}$ 
proportional to $\xi_{\mathrm{pair}}^{-2}$.

Sector E is located in the coupling region $-0.5 \lesssim \coup \lesssim + 0.6$ around unitarity and for temperature $T \lesssim T_{\rm F}$. 
It can thus be identified as a quantum degenerate unitary Fermi gas. 

\subsection{Sector F: high-temperature Boltzmann gas}
\label{sectF}

In this sector,  the rf spectra show a \emph{single\/} peak, with a width $E_{\rm w}$ following the high-temperature Boltzmann-gas benchmark (cf.~Fig.~\ref{Figure-7} and Fig.~\ref{Figure-8}(b)).
In this sector, the quasi-particle S peak (here associated to almost free fermions) is dominant over the B/QB contribution. 
This occurs even in the presence of a dispersive pole in the retarded pair propagator $\Gamma_{\rm R}$ (region to the right of the dashed line in Fig.~\ref{Figure-11}), since the kinetic energy dominates over the interaction energy in this high-temperature region.
At unitarity (or, more generally, for extremely high-temperature $T\gg  1/(m a_{\rm F}^2)$), the analytic expression $E_{\rm w} \simeq {1.25\over \sqrt{T}}$ can be obtained for the width (see the result (\ref{fwhmUL})),
which is in quite good agreement with the experimental result $E_{\rm w} \simeq {1.2 \over \sqrt{T} }$ found in Ref.~\cite{Mukherjee-2019}. 

\section{Concluding remarks}
\label{sec:conclusions}

In this article, we have provided a thorough analysis of the rf spectra of a two-component normal Fermi gas with balanced spin populations throughout the BCS-BEC crossover. 
The key discovery of the presence of a ``fixed point" in the rf spectra for different temperatures at given coupling has suggested us to analyze these spectra in terms of two underlying peaks, even when only a single peak is apparent in the spectra.

To this end, we have developed a fitting procedure that is explicitly inspired by the structure of the self-energy obtained within a $t$-matrix approach in the BEC limit.
This kind of analysis has allowed us to show that, in an appropriate region of the temperature-vs-coupling phase diagram, the size of precursor pairs in the normal phase can be extracted from the widths of the rf spectra, extending to the normal phase what was previously found in Ref.~\cite{Schunck-2008} for the size of Cooper pairs in the superfluid phase.
Interestingly, a very recent experimental work \cite{Li-2023}, has applied precisely this idea to interpret the microwave spectra of the unitary Fermi gas in the normal phase close to the critical temperature. Our analysis provides a theoretical grounding to such an extension to the normal phase of the results of  Ref.~\cite{Schunck-2008}.

More generally, we have found how the temperature dependence of the widths of the rf spectra (either of the overall single peak resulting from the sum of two peaks or of the two separate peaks, depending on the coupling strength), 
together with the relative weights of the two peaks as extracted from our fitting procedure, reveal information on the underlying physical sector in the temperature-vs-coupling phase diagram.

In addition, in appropriate temperature and coupling limits we have derived a number of analytic results for the shape and widths of the rf spectra (as detailed in the Appendices), which have been of help in extracting the relevant physical features from the rf spectra.
These analytic results have also enabled us to make a useful comparison with a fitting methodology previously introduced in Ref.~\cite{Sagi-2015} to interpret experimental momentum-resolved photoemission spectra.

A notable byproduct of the present work is the independent calculation of the pair size $\xi_{\rm pair}$ in the normal phase as a function of temperature, yielding the results presented in Fig.~\ref{Figure-9}.
This calculation was required to compare the width $E_{\rm w}$ extracted from the rf spectra with the energy scale $1/(2m \xi_{\rm pair}^2)$.
To this end, we have extracted $\xi_{\rm pair}$ from the correlation function $g_{\uparrow\downarrow}({\bf r})$ obtained following the prescriptions by Kadanoff and Baym \cite{Baym-1961,Baym-1962} 
for a diagrammatic theory to be ``conserving", which require the use of fully self-consistent Green's functions.
For the calculation of a static quantity like $g_{\uparrow\downarrow}({\bf r})$ we found it appropriate to resort to a fully self-consistent $t$-matrix approach \cite{Pini-2019, Pini-2020}, extending previous calculations based on the non-self-consistent version of the same approach \cite{Palestini-2014}.
By contrast, the rf spectra have throughout been obtained within the non-self-consistent version of the $t$-matrix approach.
This is because, as discussed in Sec.~\ref{sec:nsctheory}, it is known from many-body theory of condensed-matter systems that for dynamic quantities (like the single-particle spectral function utilized in Eq.~(\ref{irfnorm})) 
the inclusion of self-consistency for the Green's function without the simultaneous inclusion of vertex corrections may lead to incorrect physical results.

In this respect, it is relevant to point out that, although our analysis was based on a specific ($t$-matrix) approximation for the calculation of the rf spectra, the proposed fitting procedure was inspired by the structure acquired 
by the self-energy in coupling or temperature regimes where the $t$-matrix approximation is known to become asymptotically exact (i.e., in the molecular limit \cite{Haussmann-1993,Physics-Reports-2018} or at high-temperature \cite{Combescot-2006,Leyronas-2011,Parish-2013}).
For these reasons, we believe that our fitting procedure could provide a reliable scheme to interpret rf spectra under quite general conditions.

\begin{acknowledgements}
Financial support from the Italian MIUR under Project PRIN2017 (20172H2SC4) and from the European Union - NextGenerationEU through the Italian Ministry of University and Research under PNRR - M4C2 - I1.4 
Project CN00000013 ``National Centre for HPC, Big Data and Quantum Computing" is  acknowledged.
\end{acknowledgements}

\appendix   
\section{Bound state contribution to the self-energy}
\label{appA}

In this Appendix, we show how the bound-state contribution to the self-energy is computed and also provide its analytical expression in the BEC limit, justifying in this way Eqs.~(\ref{sigbecF}) and (\ref{sigbecB}) of Sec.~\ref{sec:fitting}. 
From this Appendix onwards, all expressions are given in terms of dimensionless quantities, such that wave vectors are in units of $k_{\rm F}=(3 \pi^2 n)^{1/3}$, energies and frequencies in units of $E_{\rm F}=k_{\rm F}^{2}/(2m)$, 
temperatures in units of $T_{\rm F}$, and the pair propagator $\Gamma(\mathbf{Q},\Omega)$ in units of $(2 m k_{\rm F})^{-1}$. 
Accordingly, we drop the overline to label dimensionless quantities.

Above the threshold for the formation of a two-fermion bound state (which in a many-body environment is temperature and momentum dependent, cf. Ref.~\cite{Combescot-2005}),
the pair propagator develops a pole at energy $\Omp$ below the two-body continuum threshold $\Omth={\vQ^2 \over 2}-2\mu$ where its imaginary part is infinitesimal. 
The dispersion $\Omp$ is determined by the vanishing of the real part of the inverse of $\Gamma_0^{\rm R}$ given by Eq.~(\ref{reinvg0}):
\begin{align}
-\frac{1}{8 \pi} \! \left(\frac{1}{k_{\rm F}a_{\rm F}}-\sqrt{\frac{\Omth-\Omp}{2}}\right) \! + \! \int \!\!\!\frac{d \vk}{(2 \pi)^{3}} \frac{f(\xi_{\vk+{\vQ \over 2}})}{k^{2}-\frac{\Omp-\Omth}{2}} =0 \, . 
\label{polq}
\end{align}
The numerical solution to the above equation is shown in Fig.~\ref{Figure-12} for $\coup=0.25$.
Remarkably, this solution follows quite closely the simple form $\Omp(\vQ)\simeq\alpha {\vQ^2 \over 2}-\mu_{\rm B}^*$ with $\alpha \simeq 1$ for all temperatures and couplings for which one remains above threshold. 
Here, the quantity $\mu_{\rm B}^*\equiv -\Omp(\vQ=0)$ represents an effective pair chemical potential, which reduces to the free-boson chemical potential $\mu_{\rm B}=2\mu+\be$ in the BEC limit.

Depending on coupling strength and temperature, pairs may not form for small momenta $\vQ$ due to the Pauli exclusion principle, as shown in Fig.~{\ref{Figure-12} at low temperatures. 
Nevertheless, their effective chemical potential $\mu_{\rm B}^*$ can always be obtained by extrapolation.  
Note further from Fig.~{\ref{Figure-12} that, if a pole is present for given coupling and temperature at ${\bf Q}= 0$, it then is present for all values of ${\bf Q}$. 
For this reason, the black line in Fig.~\ref{Figure-11} (full or dashed, depending on the sector), which marks the presence of a dispersive pole in the retarded pair propagator, is obtained by determining when
a solution to Eq. (\ref{polq}) at ${\bf Q} = 0$ first appears upon increasing the coupling strength at given temperature.

\begin{figure}[t!]
\begin{center}
\includegraphics[width=8.8cm]{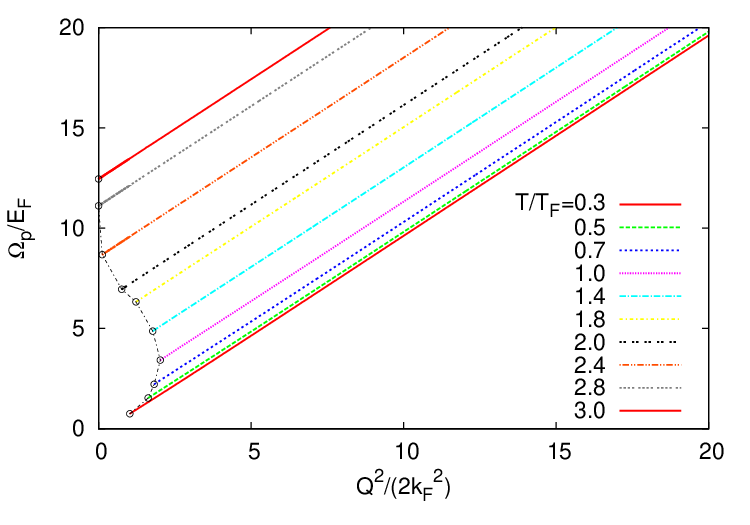}
\caption{Energy dispersion of the bound state for several temperatures when $\coup=0.25$. The line passing through the circles defines a temperature-vs-momentum region without bound states.} 
\label{Figure-12}
\end{center}
\end{figure} 

Recalling the form (\ref{sigbnd}) of the bound part of the self-energy, we rewrite it by setting $\vQ=\vk+\vp$ 
\begin{align}
\Im\left[\Sigma^{\rm bnd}(\vk, \omega)\right]=& -\int \frac{\mathrm{d} \vp}{(2 \pi)^{3}}\left[b\left(\omega+\xi_{\vp}\right)+f\left(\xi_{\vp}\right)\right] \nonumber \\
                                          & \times  \Im\left[\Gamma_{0}^{\rm polar}\left(\vk+\vp, \om+\xi_{\vp} \right)\right] \label{convkp}
\end{align}
where, aiming at performing first the integral over the angle $\theta$ between $\vk$ and $\vp$, we cast the pair spectral function in the form
\begin{align}
\Im\left[\Gamma_{0}^{\rm polar}(\vQ, \Om)\right] \Bigg\vert_{\substack{\vQ =\vk+\vp \\ \Om=\om+\xip}}  &\longrightarrow \\ \nonumber
            W_{p,\vk,\om} &\delta\left(\cos\theta-\cos\theta_{\rm p}\right)\Bigg\vert_{\substack{\vQ =\vk+\vp \\ \Om=\om+\xip}} \, .
\end{align}
Here, $\cos\theta_{\rm p}$ is defined by the solution to the equation
\beq
\Om- \Om_{\rm p}(\vQ)\Bigg\vert_{\substack{\vQ =\vk+\vp \\ \Om=\om+\xip}}=0 
\eeq
and $W_{p,\vk,\om}$ defined by
\begin{align}
W_{p,\vk,\om} &=\pi/\left|\frac{\partial\Re[\Gamma_{0}^{\rm R}(\vQ,\Om)^{-1}]}{\partial\cos\theta}\right|_{\substack{\vQ =\vk+\vp \\ \Om=\om+\xip \\ \cos\theta=\cos\theta_{\rm p} }} \nonumber \\
&=\pi /\left|2 p k\frac{\partial\Re[\Gamma_{0}^{\rm R}(\vQ,\Om)^{-1}]}{\partial \vQ^2}\right|_{\substack{\vQ =\vk+\vp \\ \Om=\om+\xip \\ \cos\theta=\cos\theta_{\rm p} }} \label{polwe} \\ 
& \equiv \frac{W}{2p k} \, .
\nonumber
\end{align}
Integration over $\cos \theta$, subject to the constraint $\left|\cos \theta_{\rm p}\right| \leq 1$, in turn defines the integration interval $\left[p^{-}_k, p^{+}_k\right]$ of the radial variable $p$, with
\beq 
	p^{\pm}_k=\left|k\pm\sqrt{2(\xik-\om-\omth)}\right|,  \label{kprim}
\eeq
where $\omth\equiv\mu_{\rm B}^*-2\mu$ (with $\omth\to\be$ in the BEC limit).

We are thus left with the expression
\begin{align}
\Im\left[\Sigma^{\rm bnd}(\vk, \om)\right]&=-\int_{{p^-_k}}^{{p^+_k}}\!\! \frac{\mathrm{d} p}{4\pi^2} W \frac{p}{2k}\Theta(\xik-\om-\omth) \nonumber\\
&\times\left[b\left(\omega+\xip\right)+f\left(\xip\right)\right] \, ,
\label{radintp}
\end{align}
where the integral can be computed analytically provided $W$ does not depend on $p$. 
This is what occurs in the BEC limit where the Fermi function in Eq.~(\ref{polq}) can be neglected, such that in Eq.~(\ref{polwe})
\beq
 \frac{\partial\Re[\Gamma_{0}^{\rm R}(\vQ,\Om)^{-1}]}{\partial \vQ^2} = \left( \frac{64\pi}{k_{\rm F}a_{\rm F}} \right)^{-1}
\eeq
yielding  $W=64\pi^2/(k_{\rm F}a_{\rm F})$. 
Away from the BEC limit, we may still continue to consider $W$ as a constant that does not depend on $p$, provided we now interpret it as a fitting parameter.
In this way, after integrating over $p$ and defining $\xi_k^\pm= (p_k^\pm)^2-\mu$, the self-energy~(\ref{convkp}) becomes
\begin{align}
&\Im \left[\Sigma^{\rm bnd}(\vk,\om)\right] =-{W \over 16\pi^2}{1 \over \beta k}\Theta(\xik-\om-\omth)\nonumber \\
             & \times\left[ \log\left(\frac{1-\mathrm{e}^{-\beta\left(\xi_k^{+}+\omega\right)}}{1-\mathrm{e}^{-\beta\left(\xi_k^{-}+\omega\right)}}\right) + 
                     \log\left(\frac{1+\mathrm{e}^{-\beta\xi_k^{-}}}{1+\mathrm{e}^{-\beta\xi_k^{+}}} \right)   \right] \, .
                     \label{asybec}
\end{align}
Note that the expressions (\ref{sigbecB}) and (\ref{sigbecF}) arise from the first and second log term within brackets in Eq.~(\ref{asybec}), respectively. 
Note also that, alhough the second log term in Eq.~(\ref{asybec}) is exponentially small in the BEC limit, it is important to keep it when using Eq.~(\ref{asybec}) as a fitting function away from the BEC limit.

\section{Comparison with an alternative \\ fitting methodology used in momentum-resolved photoemission spectroscopy }
\label{appB}

In this Appendix, we consider the momentum-resolved version of the rf spectra and compute it explicitly in the molecular (BEC) limit.
In this limit, an incoherent pairing peak, originating from the bound part of the self-energy given by Eqs.~(\ref{sigbecB}) and~(\ref{sigbecF}), clearly develops as a feature well-separated from the coherent (quasi-particle) peak
(see Sec.~\ref{sec:theoretical_approach} C-2).
Here, we show that this incoherent pairing peak, once resolved in momentum, coincides with the momentum-resolved photoemission spectrum of a thermal assembly of non-interacting molecules, previously discussed in Ref.~\cite{Sagi-2015}.

In Ref.~\cite{Sagi-2015}, the  momentum-resolved photoemission spectroscopy (PES) signal ${\cal I}(k,E)$ at momentum $k$ and energy $E$ was modeled following a Fermi liquid approach typical of solid-state systems, 
in terms of as a two-mode function:
\beq
{\cal I}(k,E)=Z {\cal I}_{\rm coh}(k,E)+(1-Z){\cal I}_{\rm inc}(k,E) \, .
\eeq
Here, the first term describes a quasi-particle state with spectral weight $Z$ and the second term corresponds to an incoherent background of collective excitations.
In the BEC limit, these two components have their rf counterparts in the S (quasi-particle) peak and B/QB (pairing) peak of the double-peak structure, respectively, that were discussed in the main text.

We are interested in the incoherent component of the spectrum, which is strictly related to the existence of atomic pairs. 
In Ref.~\cite{Sagi-2015}, the incoherent part ${\cal I}_{\rm inc}(k,E)$ of the PES signal was obtained as the dissociation spectrum of a classical (Boltzmann) gas of independent diatomic molecules with binding energy $E_{\rm p}$, 
in thermal equilibrium at temperature $T_{\rm p}$ (we use here the notation of Ref.~\cite{Sagi-2015}). 
By combining the probability for a molecule to be in a given center-of-mass momentum state with the rf line shape obtained in Ref.~\cite{Chin-2005} for the dissociation of a single weakly-bound molecule into two free atoms, 
the following analytic expression 
\begin{align}
{\cal I}_{\rm inc}(k,E)&=\Theta\left(-E_{\rm p}-E+k^{2}\right) 8\,k\,\sqrt{\frac{E_{\rm p}}{T_{\rm p}}}  \nonumber \\
                    & \times \frac{\mathrm{e}^{\frac{E_{\rm p}+E-3 k^{2}}{T_{\rm p}}} \sinh \left(\frac{2 k \sqrt{2(-E_{\rm p}-E+k^{2}})}{T_{\rm p}}\right)}{\pi^{3 / 2}\left(E-k^{2}\right)^{2}} \label{iincjin}
\end{align}
results for the incoherent part of the PES signal. 
The dimensionless  PES signal ${\cal I}(k,E)$ is normalized so that
\beq
\int_0^{\infty}\!\!\!dk\int_{-\infty}^{\infty}\!\!\!dE\; {\cal I}(k,E)=1,
\eeq
and the same normalization is used for ${\cal I}_{\rm inc}(k,E)$.

We now show that expression~(\ref{iincjin}) can be directly recovered from the incoherent part of our momentum-resolved rf spectra (PES signal) in the molecular (BEC) limit.
To this end, we recall that the normalized dimensionless PES signal ${\cal I}(k,E)$ is connected with the (dimensionless) spectral weight function $A(k,\omega)$ as follows (cf. Eq.~(\ref{irfnorm})):
\beq
{\cal I}(k,E)=3  k^{2} A(k, \om) f(\om)  \;\;\;({\rm with} \;\; \omega=E-\mu) \, .
\label{pesI}
\eeq
As mentioned above, the expression (\ref{iincjin}) was obtained in Ref.~\cite{Sagi-2015} by describing the original Fermi gas in terms of a classical gas of bosonic molecules with density $n_{\rm B}\simeq n/2$. 
Physically, this is achieved in the BEC limit $\be \gg 1$ when $T\ll \be$ for a molecular description of the gas to be valid, but at the same time when $T\gg 1$ for a Boltzmann distribution to apply to the molecular gas.  

Under these circumstances, the fermionic part~(\ref{sigbecF}) of the self-energy becomes negligible with respect to the bosonic part~(\ref{sigbecB}), since $\mu\simeq -\be/2$ and $\beta\mu\to-\infty$. 
In addition, the bosonic part takes the form 
\begin{align}
\Im \left[\Sigma_{\rm B}^{\rm bnd}(\vk,\om) \right] &=-\frac{W}{16 \pi^{2} \beta k}\Theta(\xik-\om-\omth)\nonumber\\
&\times\left[-\mathrm{e}^{-\beta\left(\xi_k^{+}+\omega\right)}+\mathrm{e}^{-\beta\left(\xi_k^{-}+\omega\right)}\right] 
\label{imsigBbec}
\end{align}
with $W ={64 \pi^2 \over k_{\rm F}a_{\rm F}}$. 
To obtain this expression, in Eq.~(\ref{sigbecB}) we have considered $\mathrm{e}^{-\beta\left(\xi_k^{\pm}+\omega\right)}\ll 1$ consistently with the assumption $1\ll T \ll \be$. 
Although in this limit we may also take $\omth \to \be$ in the result (\ref{imsigBbec}), for later use we prefer to keep the more general expression (\ref{imsigBbec}).

The remaining unbound part of the self-energy~(\ref{sigubn}) is thermally suppressed and can then be neglected.
In addition, since $|\om -\xi_k | > \omth \simeq \be$ when $\Im \left[ \Sigma_{\rm B}^{\rm bnd}(k,\om) \right] \neq 0$, it follows that in the molecular limit $\om - \xi_k$ is dominant 
with respect to the self-energy in the denominator of Eq.~(\ref{akwdef}), such that the incoherent part of the spectral function reduces to 
\beq
A_{\rm inc}(k, \om)=-\frac{1}{\pi} \frac{\Im \Sigma_{\rm B}^{\rm bnd}(k, \om)}{\left(\om-\xik\right)^{2}} \, . 
\label{akwB}
\eeq
The Fermi function $f(\om)$ in Eq.~(\ref{pesI}) can also be ignored since $\om<\xik-\omth \simeq k^2-{\be \over 2}$, so that $f(\omega)\neq 1$ only for $k^2\gtrsim \be \gg 1$. 
In this way we obtain: 
\begin{align}
&{\cal I}_{\rm inc}(k,E) = 3\,k^{2} A(k, \om=E-\mu)\nonumber \\ 
	   &= 3 k^{2} \left(-\frac{1}{\pi}\right) \frac{ -\frac{4}{ k_{\rm F}a_{\rm F} \beta k}\left[-\mathrm{e}^{-\beta\left(\xi_k^{+}+E-\mu\right)}+\mathrm{e}^{-\beta\left(\xi_k^{-}+E-\mu\right)}\right]}{(E-k^{2})^2} \, .    
\label{pesIb} 
\end{align}
By using the result (cf. Eq.~(\ref{kprim}))
\begin{align}
&\xi_k^{\pm}+E-\mu= (p_k^{\pm})^2-2\mu + E \nonumber \\
                &= 3 k^{2}-\omth-E -\mu_{\rm B}^*\pm 2 k \sqrt{2(k^{2}- E -\omth)},
\end{align}
and replacing $\omth$ and $\mu_{\rm B}^*$ respectively by $\be$ and $\mu_{\rm B}$ in the BEC  limit, after a simple algebra the expression~(\ref{pesIb}) becomes:
\begin{align}
{\cal I}_{\rm inc}(k,E)&= \Theta\left( k^{2}-E-\be \right)T \mathrm{e}^{\frac{\mu_{\rm B}}{T}} 24\,k\,\sqrt{\frac{\be}{2}}  \nonumber \\
           & \times \frac{\mathrm{e}^{\frac{\be+E-3 k^{2}}{T}} \sinh \left(\frac{2 k \sqrt{2(k^{2}-E-\be})}{T}\right)   }{\pi (E-k^{2})^2}. \label{pesmub}
\end{align}
Finally, by solving the number equation (\ref{densequM}) in the limit $1\ll T \ll \be$, where it reduces to (cf.~Eq.~(68) of Ref.~[\onlinecite{Physics-Reports-2018}])
\begin{equation}
\frac{n}{2}= \int \frac{d\bf Q}{(2\pi)^3}\mathrm{e}^{\beta(\mu_{\rm B}-Q^2/2)},
\end{equation} 
and recalling the expression $n=1/3\pi^2$ for the dimensionless density,  we get $\mathrm{e}^{\frac{\mu_{\rm B}}{T}}=\frac{1}{3} \sqrt{\frac{2}{\pi}} T^{-\frac{3}{2}}$.
In this way, we eventually obtain the expression
\begin{align}
{\cal I}_{\rm inc}(k,E)&=\Theta\left( k^{2}-E-\be \right) 8 k \sqrt{\frac{\be}{T}} \nonumber \\
               & \times \frac{\mathrm{e}^{\frac{\be+E-3 k^{2}}{T}} \sinh \left(\frac{2 k \sqrt{2(k^{2}-E-\be)}}{T}\right) }{\pi^{\frac{3}{2}}(E-k^{2})^2} \, .
\label{pesIjin}
\end{align}
This has exactly the form~(\ref{iincjin}) reported in Ref.~\cite{Sagi-2015}, with the identification $\be \rightarrow E_{\rm p}$ and $T\rightarrow T_{\rm p}$.

\section{Shape and width of the ``pairing peak"  at high temperature in the BEC region} 
\label{appCwhole}

In this Appendix, we are interested in the shape of the ``pairing peak" of $I_{\rm rf}(\omrf)$ at high temperature, in the  coupling range $a_{\rm F} > 0$ and $\be \gg 1$ where the system at low temperature would be assimilated to a BEC of molecules. 
Two different cases need be considered.

\subsection {$1\ll T \ll \be$}
\label{appC1}

This regime corresponds to that considered in Appendix~\ref{appB} for the momentum-resolved case.
In this regime, the self-energy and the incoherent part of the spectral function are given by Eqs.~(\ref{imsigBbec}) and ~(\ref{akwB}), respectively. 
Replacing $\om=\xik-\omrf$ therein, after some algebra the rf intensity~(\ref{irfnorm}) becomes
\begin{align}
& I_{\rm rf}(\omrf) = \Theta(\omrf-\omth)   \frac{3}{\omega_\delta^2} \left({4 T \over \pi k_{\rm F}a_{\rm F}}\right)  \mathrm{e}^{\beta\mu_{\rm B}^*} \nonumber \\
&  \times \int_0^\infty  \!\!\!  \!dk \,  k  \left[\mathrm{e}^{-{\left(\sqrt{2}k-\sqrt{\omrf-\omth}\right)^2 \over T }}-\mathrm{e}^{-{\left(\sqrt{2}k+\sqrt{\omrf-\omth}\right)^2 \over T }}\right] \, ,
\label{irfB}
\end{align}
where we have dropped the Fermi function as we did when deriving Eq.~(\ref{pesIb}). 
The integral over $k$ can be readily computed, yielding
\beq
 I_{\rm rf}(\omrf) = {2 \over \pi}  {\sqrt{2} \over k_{\rm F}a_{\rm F}}  \Theta(\omrf-\be)   \frac{ \sqrt{\omrf-\be}  }{ \omrf^2}
 \label{rf-Chin}
\eeq
where we have utilized the results $\omth \to \be$, $\mu_{\rm B}^* \to \mu_{\rm B}$, and $\mathrm{e}^{\beta\mu_{\rm B}}=\frac{1}{3} \sqrt{\frac{2}{\pi}} T^{-\frac{3}{2}}$.

The result (\ref{rf-Chin}) coincides with the excitation spectrum obtained in Ref.~\cite{Chin-2005} for the bound-free transition of a stationary molecule when the final-state interaction can be neglected. 
From this result one obtains the value $E_{\rm w}=1.89 \, \be$ of the FWHM. 
Recall that the same result is recovered from the mean-field (BCS) theory in the BEC limit at zero temperature (see, e.g., Ref.~\cite{Haussmann-2009}).
This is because as long as $T\ll \be$ all molecules remain bound, and their thermal motion is irrelevant in determining the rf intensity.  

\subsection{$1\ll \be  \ll T$} 
\label{appC2}

In this regime one has $|\mu| \gg T \gg \be $. 
Accordingly, for the pairing peak $\omrf$ is of the order of $\be \ll |\mu|$.
One thus has $\beta(\xik-\omrf) \gg 1$ and the Fermi function in Eq.~(\ref{irfnorm}) can be approximated by the Boltzmann factor $\mathrm{e}^{-\beta(\xik-\omrf)}$. 

In addition, the bosonic contribution of the bound part of the self-energy~(\ref{sigbecB}) becomes negligible with respect to the fermionic contribution (\ref{sigbecF}), the latter taking the form:
\beq
\Im[\Sigma^{\rm bnd}_{\rm F}(k,\omega)]= -\frac{4}{k_{\rm F}a_{\rm F}}{T \over k} \Theta(\xik-\om-\omth) \left[\mathrm{e}^{-\beta\xi_-}-\mathrm{e}^{-\beta\xi_+}\right] . 
\label{imsigbndF}
\eeq
The above bound part of the self-energy determines the pairing peak of the intensity $I_{\rm rf}(\omrf)$ through the single-particle spectral function 
\beq
A_{\rm bnd}(k,\om)=-{1\over \pi} \frac{\Im[\Sigma^{\rm bnd}_{\rm F}(k,\omega)]}{(\omega-\xik)^2} \label{akwbecT} \, ,
\eeq
where, to leading order, in the denominator we have neglected $\Sigma(k,\omega)$ with respect to $|\omega-\xi_k | \ge \omth \simeq \be \gg 1$. 
[Specifically, it can be shown that in the pairing peak region $\Sigma(k,\omega)$ is smaller at least by a factor $1/(T\sqrt{\be})$ with respect to $|\omega-\xi_k |$.]

With the result (\ref{akwbecT}) for the single-particle spectral function, we are left to evaluate the following expression:
\begin{align}
&I_{\rm rf}(\omrf) = \Theta(\omrf-\omth)   \frac{3 \mathrm{e}^{\beta\omrf}}{\omega_\delta^2} \left({4 T \mathrm{e}^{2\beta \mu}\over \pi k_{\rm F}a_{\rm F}}\right) \int_0^\infty  \!\!\!  \!dk \,  k \mathrm{e}^{-\beta k^2} \,  \nonumber \\
& \times \left[\mathrm{e}^{-\beta\left(k-\sqrt{2(\omrf-\omth)}\right)^2}-\mathrm{e}^{-\beta\left(k+\sqrt{2(\omrf-\omth)}\right)^2}\right] \, .
\label{intbecTf}
\end{align}
After integrating over $k$, this leads to
\begin{align}
I_{\rm rf}(\omrf) &=  {6  T^{3\over 2}  \mathrm{e}^{2\beta \mu} \over \sqrt{\pi} k_{\rm F}a_{\rm F}}  \mathrm{e}^{\beta\omth} \Theta(\omrf-\omth) \frac{ \sqrt{\omrf-\omth} }{ \omrf^2}\\
&= {32  \mathrm{e}^{\beta\omth}  \over 3 k_{\rm F}a_{\rm F}(\pi\,T)^{3\over 2}}  \Theta(\omrf-\omth) \frac{ \sqrt{\omrf-\omth} }{ \omrf^2} \, ,
 \label{irfhighTf}
\end{align}
where in the last line we have used the high-temperature expression $e^{\beta\mu}=\frac{4}{3\sqrt{\pi}} T^{-3/2}$ (and one could further approximate $\mathrm{e}^{\beta\omth}\simeq 1$ since $\omth\simeq \be \ll T$). 

Note that, on the BEC side of the crossover, the overall shape (\ref{irfhighTf}) of the pairing peak at high temperature coincides with that obtained at low temperature 
 (namely, for $T \ll \be$, cf.~Eq.~(\ref{rf-Chin})), when the system can be effectively described as a gas of non-interacting molecules. 
This is because at high temperature only a small fraction of molecules remain bound (as the weight proportional to $T^{-3/2}$ in Eq.~(\ref{irfhighTf}) shows), but the kinetic energy associated with the center-of-mass motion of these molecules is large and dominates the interaction effects. 
The shape of the rf spectra originating from these molecules thus coincides with that in vacuum.  
In particular, since when $\beta |\mu| \gg 1$ one has $\omth \simeq \be$, the same value $E_{\rm w} = 1.89 \be$ obtained from Eq.~(\ref{rf-Chin}) is recovered also when $T \gg \be$.

It is worth to inquire further on how the above value $E_{\rm w} = 1.89 \be$ is approached for increasing temperature. 
To this end, we need to take into account two types of effects, originating from 
(i) the temperature dependence of $\omth$ and
(ii) the corrections introduced by the self-energy which we have neglected in the denominator of the expression (\ref{akwbecT}) for the spectral weight function. 
We will show that the contributions due to these two effects are of the same order of magnitude but have opposite sign.

We first estimate the contribution of the effect (i).
Recalling that $\omth \equiv \mu_{\rm B}^*-2\mu$ as well as the definition $\mu_{\rm B}^*\equiv -\Omp(\vQ=0)$ (see Appendix~\ref{appA}), $\omth$ is seen to satisfy the equation:
\beq
-\frac{1}{8 \pi}\left(\frac{1}{k_{\rm F}a_{\rm F}}-\sqrt{\omth\over 2}\right)+ \int \frac{d \vk}{(2 \pi)^{3}} \frac{f(\xik)}{k^{2}+\frac{\omth}{2}} =0 \label{polq0} \, .
\eeq
\begin{figure}[t]
\begin{center}
\includegraphics[width=8.8cm,angle=0]{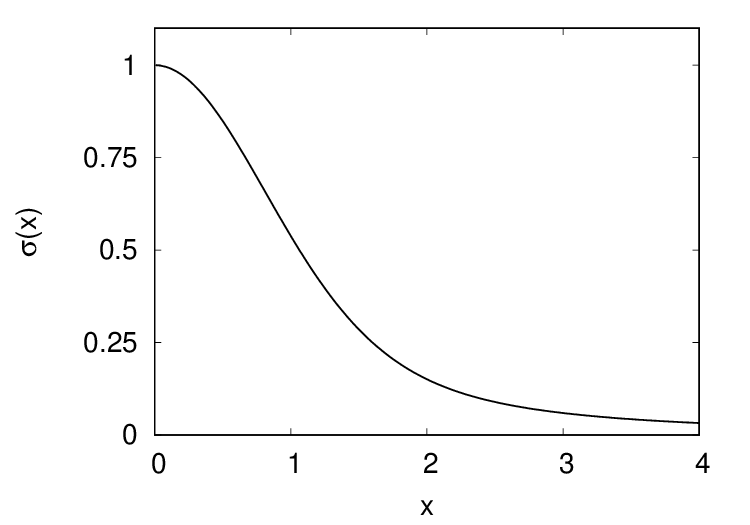}
\caption{Function $\sigma(x)\equiv D_+(x)/x$, where $D_+$ is the Dawson function \cite{Dawson}.
\color{black}}
\label{Figure-13}
\end{center}
\end{figure} 
For sufficiently high temperature when $\beta |\mu| \gg 1$, the integral in Eq.~(\ref{polq0}) can be neglected yielding $\omth=2/(k_{\rm F}a_{\rm F})^2=\be$, as anticipated above. 
The next correction to $\omth$ is obtained by evaluating the integral in Eq.~(\ref{polq0}) by replacing $f(\xi_k)$ with $e^{-\beta\xi_k}$, neglecting $\omth \ll T$ in the denominator, and using the high-temperature expression 
$e^{\beta\mu}=\frac{4}{3\sqrt{\pi}} T^{-3/2}$, yielding
\beq
\int \frac{d \vk}{(2 \pi)^{3}} \frac{f(\xik)}{k^{2}+\frac{\omth}{2}} \simeq \frac{1}{3\pi^2}\frac{1}{T} .
\eeq
With this correction, and neglecting still higher-order terms, Eq.~(\ref{polq0})  gives 
\beq
\omth=\be-{32 \over 3 \pi} {1 \over k_{\rm F}a_{\rm F}}\frac{1}{T} \, ,
\label{omthT}
\eeq
implying that, to the leading order at high temperature, $\omth$ approaches $\be$ like $T^{-1}$.

To estimate the contribution of the effect (ii), we need to evaluate $\Re[\Sigma(k,\xik-\omrf)]$ in the region of the pairing peak, when $\omrf$ is of the order of $\be$. 
In this region, we have numerically verified that the contribution to $\Re[\Sigma(k,\xik-\omrf)]$ obtained via the Kramers-Kroning transform of $\Im\Sigma^{\rm unb}$ is negligible with respect to that obtained from $\Im\Sigma^{\rm bnd}$ itself. 
We thus have
\begin{align}
&\Re[\Sigma(k,\xik-\omrf)] = \frac{1}{\pi} {\cal P}\!\!\int \!\!d\om \frac{\Im \Sigma^{\rm bnd}_{\rm F}(k,\om)}{\om-\xik+\omrf} \label{eqRe1}\\
&=\frac{8e^{\beta\mu}}{\pi k_{\rm F}a_{\rm F}}\frac{T}{k}{\cal P}\!\!\int_0^\infty \!\!dy \,y \,\frac{e^{-\beta(k+y)^2}-e^{-\beta(k-y)^2}}{y^2-2(\omrf-\omth)}
\label{eqRe2}\\
&=\frac{16e^{\beta\mu}}{\pi k_{\rm F}a_{\rm F}}\frac{\sqrt{T}}{\tilde{k}}e^{-\tilde{k}^2}{\cal P}\!\!\int_0^\infty \!\!d\tilde{y}\,\tilde{y} \frac{e^{-\tilde{y}^2} \sinh(2\tilde{k}\tilde{y})}{\tilde{y}^2-2\beta(\omrf-\omth)} \label{eqRe3},
\end{align}
where we have set $\om=\xik-\omth-y^2/2$ from line (\ref{eqRe1}) to line (\ref{eqRe2}), as well as $y=\sqrt{T} \tilde{y}$ from line (\ref{eqRe2}) to line (\ref{eqRe3}), and we have further defined $\tilde{k}\equiv k/\sqrt{T}$.
In the denominator of the expression (\ref{eqRe3}) we can neglect $2\beta(\omrf-\omth)$ because, in the region of the peak, $\omrf-\omth=O(\be)\ll T$. 
Integration over $\tilde{y}$ then yields
\begin{align}
\Re[\Sigma(k,\xik-\omrf)] &=\frac{16e^{\beta\mu}\sqrt{T}}{\sqrt{\pi} k_{\rm F}a_{\rm F}}\frac{1}{\tilde{k}}D_+(\tilde{k})\\
&=\frac{64}{3\pi k_{\rm F}a_{\rm F}}\frac{\sigma(\tilde{k})}{T} 
\end{align}
where $\sigma(\tilde{k})\equiv D_+(\tilde{k})/\tilde{k}$ with $D_+$ the Dawson function \cite{Dawson}. 
A plot of the function $\sigma(x)$ is shown in Fig.~\ref{Figure-13} for convenience.
With this result for $\Re[\Sigma(k,\xik-\omrf)]$, the integral~(\ref{intbecTf}) is now modified to read:
\begin{align}
&I_{\rm rf}(\omrf) = \Theta(\omrf-\omth) \left({12 T \mathrm{e}^{2\beta \mu}\mathrm{e}^{\beta\omrf}\over \pi k_{\rm F}a_{\rm F}}\right) \int_0^\infty  \!\!\!  \!dk \,  k \mathrm{e}^{-\beta k^2} \,  \nonumber \\
& \times \frac{\mathrm{e}^{-\beta\left(k-\sqrt{2(\omrf-\omth)}\right)^2}-\mathrm{e}^{-\beta\left(k+\sqrt{2(\omrf-\omth)}\right)^2}}{ \left( \omrf+ \frac{64}{3\pi k_{\rm F}a_{\rm F}} {\sigma\left({k/\sqrt{T}}\right) \over T} \right)^2 } \, .
\label{intbecTf_2}
\end{align}
In the denominator of (\ref{intbecTf_2}) we have neglected the term originating from $\Im[\Sigma^{\rm bnd}_{\rm F}(k,\xik-\omrf)]$ because it is smaller by a factor $\be/T$ with respect to the smallest term we have retained 
(which originates from $\Re[\Sigma(k,\xik-\omrf)]$). 

The term containing $\sigma({k/\sqrt{T}})$ in the denominator of Eq.~(\ref{intbecTf_2}) is small compared to $\omrf \ge \omth$ by a factor $1/(\sqrt{\be} T)$, so that we can expand the expression (\ref{intbecTf_2}) and obtain 
a correction proportional to $1/(\omrf)^3$.
We can thus replace $\sigma({k/\sqrt{T}})$ in the denominator of (\ref{intbecTf}) by a constant value $\sigma_0$, by requiring that the coefficient of the term proportional to $1/(\omrf)^3$ coincides 
with that obtained by keeping the full expression for $\sigma\left({k/\sqrt{T}}\right)$.  
This requirement leads us to the result
\beq
\sigma_0=\frac{\int_0^\infty \!\!d\tilde{k} \,\tilde{k} \,\mathrm{e}^{-2\tilde{k}^2}\sinh\left(2\tilde{k}\sqrt{2\beta(\omrf-\omth)}\right)\sigma(\tilde{k})}{\int_0^\infty \!\!d\tilde{k} \,\tilde{k} \,\mathrm{e}^{-2\tilde{k}^2}\sinh\left(2\tilde{k}\sqrt{2\beta(\omrf-\omth)}\right)} \, .
\eeq
In this expression we can further approximate $\sinh(x)\simeq x$ since $\beta(\omrf-\omth)\ll 1$ and $\tilde{k}=O(1)$, yielding
\beq
\sigma_0=\frac{\int_0^\infty \!\!d\tilde{k} \,\tilde{k}^2 \,\mathrm{e}^{-2\tilde{k}^2}\sigma(\tilde{k})}{\int_0^\infty \!\!d\tilde{k} \,\tilde{k}^2 \,\mathrm{e}^{-2\tilde{k}^2}}=\frac{\sqrt{\pi/2}/12}{\sqrt{\pi/2}/8}=\frac{2}{3} \, .
\eeq
In this way, we obtain for the temperature correction to expression~(\ref{irfhighTf}):
\begin{align}
I_{\rm rf}(\omrf) = {32  \mathrm{e}^{\beta\omth}  \over 3 k_{\rm F}a_{\rm F}(\pi\,T)^{3\over 2}}  \Theta(\omrf-\omth) \frac{ \sqrt{\omrf-\omth} }{\left(\omrf+{s_0 \over T}\right)^2}  
\end{align}
where $s_0=128/(9\pi k_{\rm F}a_{\rm F})$.
With a simple change of variable, the resulting FWHM turns out to be 
\beq
E_{\rm w}=1.89\left(\omth+{ s_0\over T}\right),
\label{less-asy}
\eeq
which using the expansion (\ref{omthT}) for $\omth$ becomes eventually:
\begin{align}
E_{\rm w} &\simeq 1.89 \left(\be + {32 \over 9\pi k_{\rm F}a_{\rm F}} {1\over T}\right).\label{asymBEC}
\end{align}

\begin{figure}[t]
\begin{center}
\includegraphics[width=8.8cm,angle=0]{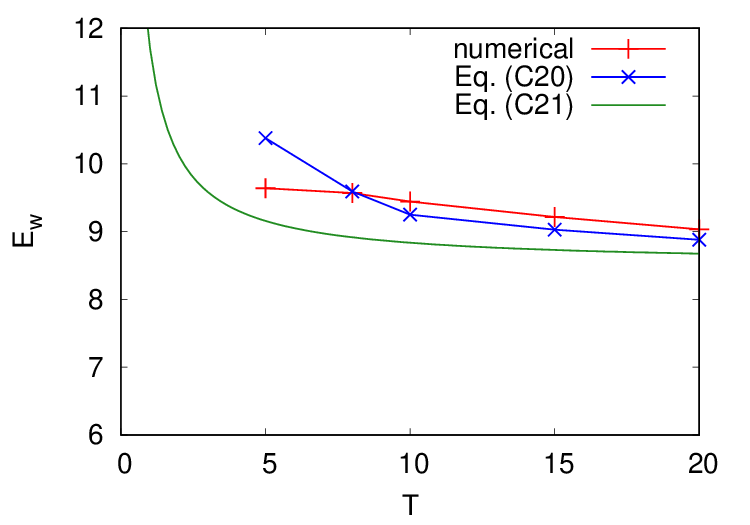}
\caption{Comparison between the full numerical calculation of $E_{\rm w}$ for the pairing peak when $\coup=1.5$ (full line) and the asymptotic expressions (\ref{asymBEC}) and (\ref{less-asy}) (dashed line and dashed-dotted line, respectively).}
\label{Figure-14}
\end{center}
\end{figure}

The comparison between the high-temperature expansion (\ref{asymBEC}) and the full numerical calculation of $E_{\rm w}$ for the pairing peak when $\coup=1.5$ is shown in Fig.~\ref{Figure-14}. 
One sees that the numerical $E_{\rm w}$ approaches the asymptotic expression (\ref{asymBEC}), albeit rather slowly. 
The reason for this slow approach resides in the approximation (\ref{omthT}) for $\omth$. 
The next-order correction to Eq.~(\ref{omthT}) yield the term ${32 \over 3 \sqrt{\pi}} {1 \over (k_{\rm F}a_{\rm F})^2}\frac{1}{T^{3/2}}$ which is still significant even at the highest temperature ($T=20$) considered in Fig.~\ref{Figure-14}. 
On the other hand, the less asymptotic expression (\ref{less-asy}), where $\omth$ is obtained by Eq.~(\ref{polq0}) with $f(\xi_k)\simeq \frac{4}{3\sqrt{\pi}T^{3/2}} e^{-\beta k^2}$, reproduces quite well the numerical 
$E_{\rm w}$ already at $T\simeq 10$.

\section{Width of rf spectra in the Boltzmann limit when $T\gg \be$ or $1\ll T \ll \be$ and  $a_{\rm F} < 0$}
\label{appD}

In this Appendix, we derive the FWHM of the rf spectra in the Boltzmann (high-temperature) limit ($T\gg 1$, such that $z=e^{\beta \mu} \ll 1$) in two different regimes: (i) $T\gg \be$ and (ii) $1 \ll T \ll \be$ with $a_{\rm F} < 0$.

Quite generally, statistical effects of the medium on two-particle interactions become negligible in the Boltzmann limit. 
One is thus allowed to approximate the $t$-matrix by its counterpart in vacuum given by the expression (\ref{2btm}), yielding:
\beq
\Im\left[\Gamma_{0}^{R}(\vQ, \Om)\right]=\frac{8 \pi \sqrt{2} \sqrt{\Omega-\frac{Q^{2}}{2}+2 \mu}}{\be+\Omega-\frac{Q^{2}}{2}+2 \mu} .\label{img0vac}
\eeq
In addition, in both the regimes (i) and (ii) that we are considering, the single-particle spectral function is dominated by the quasi-particle peak located about $\omega=\xi_k$, since at high temperature the self-energy provides a small correction 
to the free-particle dispersion. 
The imaginary part of the self-energy given by Eq.~(\ref{imsancon}) at $\omega=\xik$ then acquires the form
\begin{align}
\Im \left[\Sigma\left(\vk,\xik\right)\right] = 
    -\int\!\! \frac{d \vQ}{(2 \pi)^{3}} \frac{16 \pi Q e^{-\beta \xi_{\vQ+\vk}}}{2 \be + Q^2 } \, ,
\end{align}
where the term containing the Bose function in Eq.~(\ref{imsancon}) has been neglected being smaller by a factor $z$ than the term with the Fermi function.
In addition, the Fermi function, has been replaced by a Boltzmann factor and the change of variable $\vQ \rightarrow \vQ+2\vk$ has been performed.
The angular integration can then be done analytically, yielding
\beq
\Im \left[\Sigma\left(k,\xik\right)\right] =-\frac{16 e^{-\tilde{k}^2}}{3\pi^{3\over 2} k}\int_0^\infty \!\!\! d\tilde{Q}\frac{\tilde{Q}^2 e^{-\tilde{Q}^2}\sinh(2\tilde{k}\tilde{Q})}{2 \tilde{\varepsilon}_0 + \tilde{Q}^2},
\label{eqhigT_a_neg}
\eeq
where we have defined $\tilde{Q}\equiv Q/\sqrt{T}$, $\tilde{k}=k/\sqrt{T}$, $\tilde{\varepsilon}_0=\be/T$, and used $e^{\beta \mu} T^{3 \over 2}=\frac{4}{3 \sqrt{\pi}}$.
Similarly to Appendix~\ref{appCwhole}, two different cases need be considered at this point.
 
\begin{figure}[t]
\begin{center}
\includegraphics[width=8.8cm,angle=0]{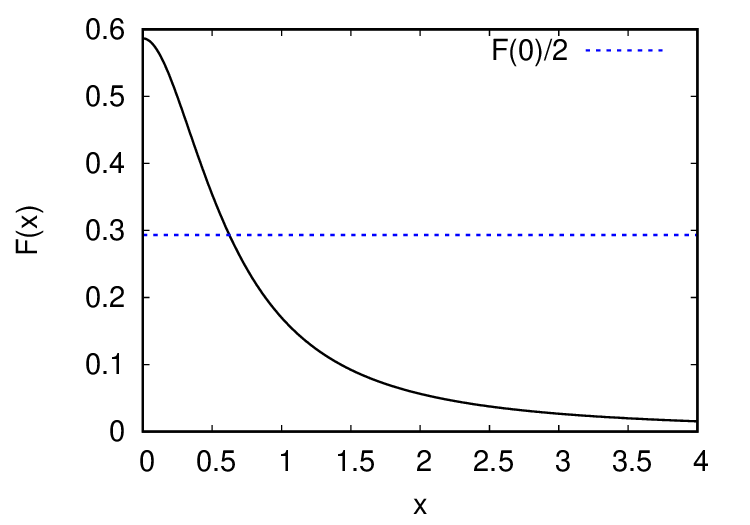}
\caption{Function $F(x)$ as defined by Eq.~(\ref{Fofx}).}
\label{Figure-15}
\end{center}
\end{figure} 

\subsection{ $T \gg \be$}
\label{appD1}

When $T\gg \be$,  $\tilde{\varepsilon}_0$ can be neglected in Eq.~(\ref{eqhigT_a_neg}), yielding
\begin{align}
\Im \left[\Sigma\left(k,\xik\right)\right] &=-\frac{16 e^{-\tilde{k}^2}}{3\pi^{3\over 2} k}\int_0^\infty \!\!\! d\tilde{Q}e^{-\tilde{Q}^2}\sinh(2\tilde{k}\tilde{Q})\\
&=-\frac{8}{3 \pi} \frac{\operatorname{erf}(k / \sqrt{T})}{k} \equiv-\gamma_k \, ,
\label{sigUL}
\end{align}
a result that was first obtained in Ref.~\cite{Enss-2011} for the unitary Fermi gas (for which $\be=0$).
The form (\ref{sigUL}) can be utilized for the single-particle spectral function entering the expression (\ref{irfnorm}) of the rf intensity, giving
\beq
\Irf(\omrf)=3 \int_0^{+\infty} \!\!\! \mathrm{d} k\, k^2\,  \frac{1}{\pi} \frac{\gamma_{k}}{\om_{\delta}^{2}+\gamma_{k}^2} e^{-\frac{k^2}{T}} e^{\beta \mu} \, . 
\label{irfboltz}
\eeq
To obtain this result for $\Irf(\omrf)$, we have: 
(a) approximated $\Im \left[\Sigma\left(k,\xik-\omrf\right)\right]\simeq\Im \left[\Sigma\left(k,\xik\right)\right]$; 
(b) neglected $\Re \left[\Sigma\left(k,\xik-\omrf\right)\right]$ 
because it is sub-leading with respect to $\Im \left[\Sigma\left(k,\xik-\omrf\right)\right]$ as far as the width of the peak is concerned; 
(c) set $e^{\beta\omrf}=1$ in the Boltzmann factor since around the peak $\omrf$ is of order $\gamma_k \ll T$.
Changing further the integration variable to $\tilde{k}=k / \sqrt{T}$ and using $e^{\beta \mu} T^{3/2}=\frac{4}{3 \sqrt{\pi}}$, we obtain eventually
\beq
\Irf(\omrf)=\frac{3}{2}\sqrt{T\over \pi}  F(\sqrt{T}\omrf)
\eeq
with
\beq
F(x)=\int_{0}^{+\infty}\!\!\! \frac{\operatorname{erf}(\tilde{k}) /\tilde{k}}{(3 \pi x/8)^2 +\operatorname{erf}(\tilde{k})^{2} / \tilde{k}^{2}} \tilde{k}^{2} e^{-\tilde{k}^{2}} d \tilde{k} \, .
\label{Fofx}
\eeq

Numerical integration of this expression yields the plot of $F(x)$ shown in Fig.~\ref{Figure-15}. 
From this plot one gets that the half-peak value $F(0)/2$ is obtained for $|x| \simeq 0.624$, which corresponds to the FWHM 
\beq
E_{\rm w}={2 \times 0.624 \over \sqrt{T}} \simeq {1.25 \over \sqrt{T}}
\label{fwhmUL}
\eeq
for the peak of $\Irf(\omrf)$.
This result  is in quite good agreement with the experimental behavior ${1.2 \over \sqrt{T}}$ found in Ref.~\cite{Mukherjee-2019} at unitarity, and is consistent with the relaxation time $\tau \propto T^{1/2}$ previously found in 
Refs.~\cite{Bruun-2007,Enss-2011}.

 \begin{figure}[t]
\begin{center}
\includegraphics[width=8.8cm,angle=0]{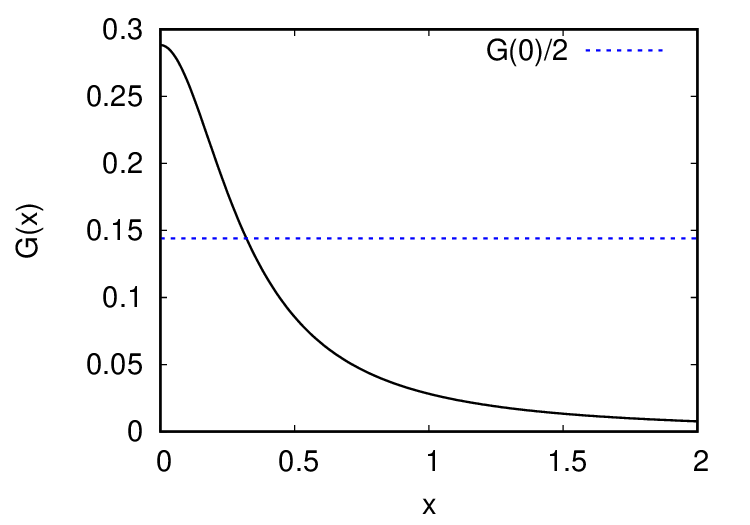}
\caption{Function $G(x)$ as defined by Eq.~(\ref{Gofx}).}
\label{Figure-16}
\end{center}
\end{figure}


\subsection{$1\ll T \ll \be$ with $a_{\rm F} < 0$}  
\label{appD2}

This condition implies a weak coupling strength $\coup \ll -1$ and a high temperature compared to $T_{\rm F}$, although still smaller than the energy scale $\be = (m a_{F}^{2})^{-1}$.
Under these circumstances, $\tilde{Q}^2$ can be neglected with respect to $\tilde{\varepsilon}_0$ in the denominator of Eq.~(\ref{eqhigT_a_neg}), resulting in
\beq
\Im\left[\Sigma\left(k, \xik\right)\right] =-\frac{2\sqrt{T}\!\left(k_{\rm F} a_{\rm F}\right)^{2}}{3 \pi}g(\tk) \label{sigWC}
\eeq
with 
\beq
g(\tk)=\left(\tk+\frac{1}{2 \tk}\right) \operatorname{erf}(\tk)+e^{-\tk^{2}}/\sqrt{\pi} \, .
\eeq
Once the self-energy (\ref{sigWC}) is inserted into the expression of the single-particle spectral function needed in Eq.~(\ref{irfnorm}) for the rf intensity (and again dropping $\Re \Sigma$), we obtain eventually:
\begin{align}
&\Irf(\omrf) ={6 \over \sqrt{\pi}} {1 \over (k_{\rm F}a_{\rm F})^2 \sqrt{T}}  G\left({\om_{\delta}\over \sqrt{T}(k_{\rm F}a_{\rm F})^2}\right)
\end{align}
with
\beq
G(x)=  \int_0^{+\infty} \hspace{-0.2cm} \mathrm{d}\tk\,  \, \frac{g(\tk)}{\left(3 \pi x/2\right)^2+g(\tk)^2}\tk^2 e^{-\tk^2}
\label{Gofx}
\eeq

Numerical integration of this expression yields the plot of $G(x)$ shown in Fig.~\ref{Figure-16}.  
From this plot one gets that the half-peak value $G(0)/2$ is obtained for $|x| \simeq 0.320$, which corresponds to the FWHM 
\beq
E_{\rm w}\simeq 0.64  (k_{\rm F}a_{\rm F})^2 \sqrt{T}
\label{fwhmsqrtT}
\eeq
for the peak of $\Irf(\omrf)$.
A somewhat analogous result was obtained in Ref.~\cite{Massignan-2005} for the viscosity relaxation time $\tau_\eta$ in the same regime  $1\ll T \ll \be$ (with $a_{\rm } < 0)$, for which 
\beq
{1\over \tau_\eta}=\frac{32\sqrt{2}}{15 \pi^{3\over 2}}(k_{\rm F}a_{\rm F})^2 \sqrt{T}\simeq 0.54 (k_{\rm F}a_{\rm F})^2 \sqrt{T} \, .
\eeq

\newpage
	


\begin{thebibliography}{99}

\bibitem{Ketterle-2008}
W. Ketterle and M.~W. Zwierlein, \emph{Making, probing and understanding ultracold Fermi gases\/}, Riv. Nuovo Cimento {\bf 31}, 247 (2008).

\bibitem{Vale-2021}
C.~Vale and M.~Zwierlein, \emph{Spectroscopic probes of quantum gases\/}, Nat. Phys. {\bf 17}, 1305 (2021).

\bibitem{Regal-2003-a} 
C.~A. Regal and D.~S. Jin, \emph{Measurement of positive and negative scattering lengths in a Fermi gas of atoms\/}, Phys. Rev. Lett. {\bf 90}, 230404 (2003).

\bibitem{Gupta-2003} 
S. Gupta, Z. Hadzibabic, M.~W. Zwierlein, C.~A. Stan, K. Dieckmann, C.~H. Schunck, E. ~G.~M. van Kempen, B.~J. Verhaar, and W. Ketterle, \emph{Radio-frequency spectroscopy of ultracold fermions\/}, Science {\bf 300}, 1723 (2003).

\bibitem{Regal-2003-b} 
C.~A. Regal, C. Ticknor, J.~L. Bohn, and D.~S. Jin, \emph{Creation of ultracold molecules from a Fermi gas of atoms\/}, Nature {\bf 424},  47 (2003).

\bibitem{Chin-2004} 
C. Chin, M. Bartenstein, A. Altmeyer, S. Riedl, S. Jochim, J. Hecker Denschlag, and R. Grimm, \emph{Observation of the pairing gap in a strongly interacting Fermi gas\/}, Science {\bf 305}, 1128 (2004).

\bibitem{Shin-2007} 
Y. Shin,  C.~H. Schunck, A. Schirotzek, and W. Ketterle, \emph{Tomographic rf spectroscopy of a trapped Fermi gas at unitarity\/}, Phys. Rev. Lett. {\bf 99}, 090403 (2007).

\bibitem{Stewart-2008} 
J.~T. Stewart, J.~P Gaebler, and  D.~S. Jin, \emph{Using photoemission spectroscopy to probe a strongly interacting Fermi gas\/}, Nature {\bf 454}, 744 (2008).

\bibitem{Schirotzek-2008} 
A. Schirotzek, Y. Shin, C.~H. Schunck, and W. Ketterle, \emph{Determination of the superfluid gap in atomic Fermi gases by quasiparticle spectroscopy\/},  Phys. Rev. Lett. {\bf 101}, 140403 (2008).

\bibitem{Schunck-2008} 
C.~H. Schunck, Y. Shin, A. Schirotzek, and W. Ketterle, \emph{Determination of the fermion pair size in a resonantly interacting superfluid\/}, Nature {\bf 454}, 739 (2008).

\bibitem{Gaebler-2010} 
J.~P. Gaebler, J.~T. Stewart, T.~E. Drake, D.~S. Jin, A. Perali, P. Pieri, and G.~C. Strinati, \emph{Observation of pseudogap behaviour in a strongly interacting Fermi gas\/}, Nat. Phys. {\bf 6}, 569 (2010).

\bibitem{Shkedrov-2018}
C. Shkedrov, Y. Florshaim, G. Ness, A. Gandman, and Y. Sagi, \emph{High-sensitivity rf spectroscopy of a strongly interacting Fermi gas\/}, Phys. Rev. Lett. {\bf 121}, 093402 (2018).

\bibitem{Mukherjee-2019}
B. Mukherjee, P.~B. Patel, Z. Yan, R.~J. Fletcher, J. Struck, and M.~W. Zwierlein, \emph{Spectral response and contact of the unitary Fermi gas\/}, Phys. Rev. Lett. {\bf 122}, 203402 (2019).

\bibitem{Kinnunen-2004} 
J. Kinnunen, M. Rodriguez, and P. T\"orm\"a, \emph{Pairing gap and in-gap excitations in trapped fermionic superfluids\/}, Science {\bf 305}, 1131 (2004).

\bibitem{He-2005} 
Y. He, Q. Chen, and K. Levin, \emph{Radio-frequency spectroscopy and the pairing gap in trapped Fermi gases\/}, Phys. Rev. A {\bf 72}, 011602 (2005).

\bibitem{Ohashi-2005} 
Y. Ohashi and A. Griffin, \emph{Single-particle excitations in a trapped gas of Fermi atoms in the BCS-BEC crossover region\/}, Phys. Rev. A {\bf 72}, 013601 (2005).

\bibitem{Yu-2006}
Z. Yu and G. Baym,  \emph{Spin-correlation functions in ultracold paired atomic-fermion systems: Sum rules, self-consistent approximations, and mean fields\/}, Phys. Rev. A {\bf 73}, 063601 (2006).

\bibitem{Punk-2007}
M. Punk and W. Zwerger, \emph{Theory of rf-spectroscopy of strongly interacting fermions\/}, Phys. Rev. Lett. {\bf 99}, 170404 (2007).

\bibitem{Baym-2007}
G. Baym, C.~J. Pethick, Z. Yu, and M.~W. Zwierlein, \emph{Coherence and clock shifts in ultracold Fermi gases with resonant interactions\/}, Phys. Rev. Lett. {\bf 99}, 190407 (2007).

\bibitem{Perali-2008}
A. Perali, P. Pieri, and G.~C. Strinati, \emph{Final-state and pairing-gap effects in the radio-frequency spectra of ultracold Fermi atoms\/}, Phys. Rev. Lett. {\bf 100}, 010402 (2008).

\bibitem{Massignan-2008}
P. Massignan, G.~M. Bruun, and H.~T.C. Stoof, \emph{Twin peaks in rf spectra of Fermi gases at unitarity\/}, Phys. Rev. A {\bf 77}, 031601 (2008). 

\bibitem{Pieri-2009}
P. Pieri, A. Perali, and G.~C. Strinati, \emph{Enhanced paraconductivity-like fluctuations in the radiofrequency spectra of ultracold Fermi atoms\/}, Nat. Phys. {\bf 5}, 736 (2009).

\bibitem{Haussmann-2009} 
R. Haussmann, M. Punk, and W. Zwerger, \emph{Spectral functions and rf response of ultracold fermionic atoms\/}, Phys. Rev. A {\bf 80}, 063612 (2009).

\bibitem{Pieri-2011}
P. Pieri, A. Perali, G.~C. Strinati, S. Riedl, M.~J. Wright, A. Altmeyer, C. Kohstall, E.~R. S\'anchez Guajardo, J. Hecker Denschlag, and R. Grimm,
\emph{Pairing-gap, pseudogap, and no-gap phases in the radio-frequency spectra of a trapped unitary $^6$Li gas\/}, Phys. Rev. A {\bf 84}, 011608 (2011).

\bibitem{Torma-2016}
P. T\"orm\"a, \emph{Physics of ultracold Fermi gases revealed by spectroscopies\/}, Phys. Scr. {\bf 91}, 043006 (2016).

\bibitem{Pistolesi-1994}
F. Pistolesi and G.~C. Strinati, \emph{Evolution from BCS superconductivity to Bose condensation: Role of the parameter $k_{\rm F} \xi$\/}, Phys. Rev. B {\bf 49}, 6356 (1994).

\bibitem{Pistolesi-1996}
F. Pistolesi and G.~C. Strinati, \emph{Evolution from BCS superconductivity to Bose condensation: Calculation of the zero-temperature phase coherence length\/}, Phys. Rev. B {\bf 53},  15168 (1996).

\bibitem{Palestini-2014}
F. Palestini and G.~C. Strinati, \emph{Temperature dependence of the pair coherence and healing lengths for a fermionic superfluid throughout the BCS-BEC crossover\/}, Phys. Rev. B {\bf 89}, 224508 (2014).

\bibitem{Tsuchiya-2010} 
S. Tsuchiya, R. Watanabe, and Y. Ohashi, \emph{Photoemission spectrum and effect of inhomogeneous pairing fluctuations in the BCS-BEC crossover regime of an ultracold Fermi gas\/}, Phys. Rev. A {\bf 82}, 033629 (2010).

\bibitem{Perali-2011}
A. Perali, F. Palestini, P. Pieri, G.~C. Strinati, J. T. Stewart, J. P. Gaebler, T. E. Drake, and D. S. Jin, \emph{Evolution of the normal state of a strongly interacting Fermi gas from a pseudogap phase to a molecular Bose gas\/},
Phys. Rev. Lett. {\bf 106}, 060402 (2011).

\bibitem{Ota-2017}
M. Ota, H. Tajima, R. Hanai, D. Inotani, and Y. Ohashi,
\emph{Local photoemission spectra and effects of spatial inhomogeneity in the BCS-BEC-crossover regime of a trapped ultracold Fermi gas\/}, Phys. Rev. A {\bf 95}, 053623 (2017).

\bibitem{Hu-2022}
H. Hu and X.-J. Liu,
\emph{Fermi polarons at finite temperature: Spectral function and rf spectroscopy\/}, Phys. Rev. A {\bf 105}, 043303 (2022).


\bibitem{Schmidt-2012}
R. Schmidt, T. Enss, V. Pietil\"a, and E. Demler,
\emph{Fermi polarons in two dimensions\/}, Phys. Rev. A {\bf 85}, 021602(R) (2012).

\bibitem{Pietila-2012}
V. Pietil\"a, \emph{Pairing and radio-frequency spectroscopy in two-dimensional Fermi gases\/}, Phys. Rev. A {\bf 86}, 023608 (2012).

\bibitem{Watanabe-2013}
R. Watanabe, S. Tsuchiya, and Y. Ohashi,
\emph{Low-dimensional pairing fluctuations and pseudogapped photoemission spectrum in a trapped two-dimensional Fermi gas\/}, Phys. Rev. A {\bf 88}, 013637 (2013).

\bibitem{Marsiglio-2015}
F. Marsiglio, P. Pieri, A. Perali,  F. Palestini, and G.~C. Strinati, \emph{Pairing effects in the normal phase of a two-dimensional Fermi gas\/}, Phys. Rev. B {\bf 91}, 054509 (2015).

\bibitem{Sagi-2015} 
Y. Sagi, T. E. Drake, R. Paudel, R. Chapurin, and D. S. Jin, \emph{Breakdown of the Fermi liquid description for strongly interacting fermions\/}, Phys. Rev. Lett. {\bf 114}, 075301 (2015).


\bibitem{Mukherjee-2017}
B. Mukherjee, Z. Yan, P. B. Patel, Z. Hadzibabic, T. Yefsah, J. Struck, and M. W. Zwierlein, \emph{Homogeneous Atomic Fermi Gases\/}, Phys. Rev. Lett. {\bf 118}, 123401 (2017). 

\bibitem{Hueck-2018}
K. Hueck, N. Luick, L. Sobirey, J. Siegl, T. Lompe, and H. Moritz, \emph{Two-Dimensional Homogeneous Fermi Gases\/}, Phys. Rev. Lett. {\bf 120}, 060402 (2018). 

\bibitem{Yan-2019}
Z. Yan, P. B. Patel, B. Mukherjee, R. J. Fletcher, J. Struck, and M. W. Zwierlein, \emph{Boiling a Unitary Fermi Liquid\/}, Phys. Rev. Lett. {\bf 122}, 093401 (2019). 

\bibitem{Shkedrov-2022}
C. Shkedrov, M. Menashes, G. Ness, A. Vainbaum, and Y. Sagi, \emph{Absence of heating in a uniform Fermi gas created by periodic driving}, Phys. Rev. X {\bf 12}, 011041 (2022).

\bibitem{Sademelo-1993}
C.~A.~R. S\'a de Melo, M. Randeria, and J.~R. Engelbrecht, \emph{Crossover from BCS to Bose superconductivity: Transition temperature and time-dependent Ginzburg-Landau theory\/}, Phys. Rev. Lett. {\bf 71}, 3202 (1993).

\bibitem{Pieri-2000} 
P. Pieri and G.~C. Strinati, \emph{Strong-coupling limit in the evolution from BCS superconductivity to Bose-Einstein condensation\/}, Phys. Rev. B {\bf 61}, 15370 (2000).

\bibitem{Simonucci-2005}
S. Simonucci, P. Pieri, and G.~C. Strinati, \emph{Broad vs narrow Fano-Feshbach resonances in the BCS-BEC crossover with trapped Fermi atoms\/}, Europhys. Lett. {\bf 69}, 713 (2005).

\bibitem{Haussmann-1993} R. Haussmann,  \emph{Crossover from BCS superconductivity to Bose-Einstein condensation: A self-consistent theory\/}, Z. Phys. B {\bf 91}, 291 (1993).

\bibitem{Haussmann-1994} R. Haussmann, \emph{Properties of a Fermi liquid at the superfluid transition in the crossover region between BCS superconductivity and Bose-Einstein condensation\/}, Phys. Rev. B {\bf 49}, 12975 (1994).

\bibitem{Micnas-1995} R. Micnas, M.~H. Pedersen,  S. Schafroth, T. Schneider, J. J. Rodr\'iguez-N\'u\~{n}ez, and H. Beck, \emph{Excitation spectrum of the attractive Hubbard model\/}, Phys. Rev. B {\bf 52}, 16223  (1995).

\bibitem{Yanase-1999} Y. Yanase and K. Yamada, \emph{Theory of pseudogap phenomena in High-T$_c$ cuprates based on the strong coupling superconductivity\/}, J. Phys. Soc. Jpn. {\bf 68}, 2999 (1999).

\bibitem{Rohe-2001}
D. Rohe and W. Metzner, \emph{Pair-fluctuation-induced pseudogap in the normal phase of the two-dimensional attractive
Hubbard model at weak coupling\/}, Phys. Rev. B {\bf 63}, 224509 (2001).

\bibitem{Perali-2002}
A. Perali, P. Pieri, C. Castellani, and G.~C. Strinati, \emph{Pseudogap and spectral function from superconducting fluctuations to the bosonic limit\/}, Phys. Rev. B {\bf 66}, 024510 (2002).

\bibitem{Pieri-2004}
P. Pieri, L. Pisani, and G.~C. Strinati, \emph{BCS-BEC crossover at finite temperature in the broken-symmetry phase\/}, Phys. Rev. B {\bf 70}, 094508 (2004).

\bibitem{Nozieres-1985}
P. Nozi\`eres and S. Schmitt-Rink,
\emph{Bose condensation in an attractive fermion gas: From weak to strong coupling superconductivity\/}, J. Low Temp. Phys. {\bf 59},195 (1985). 

\bibitem{Serene-1989}
J.~W. Serene, \emph{Stability of two-dimensional Fermi liquids against pair fluctuations with large total momentum\/}, 
Phys. Rev. B {\bf 40}, 10873 (1989).

\bibitem{Thouless-1960}
D.~J. Thouless, \emph{Perturbation theory in statistical mechanics and the theory of superconductivity\/}, Ann. Phys.  {\bf 10}, 553 (1960).

\bibitem{Physics-Reports-2018} 
G.~C. Strinati, P. Pieri, G. R\"{o}pke, P. Schuck, and M. Urban, \emph{The BCS-BEC crossover: From ultra-cold Fermi gases to nuclear systems\/}, Phys. Rep. {\bf 738}, 1 (2018).

\bibitem{Combescot-2006} 
R. Combescot, X. Leyronas, and M. Yu. Kagan,
\emph{Self-consistent theory for molecular instabilities in a normal degenerate Fermi gas in the BEC-BCS crossover\/},
Phys. Rev. A {\bf 73}, 023618 (2006).

\bibitem{Liu-2009}
X.-J. Liu, H. Hu, and P.~D. Drummond, \emph{Virial Expansion for a Strongly Correlated Fermi Gas\/},
Phys. Rev. Lett. {\bf 102}, 160401 (2009).

\bibitem{Leyronas-2011} 
X. Leyronas, \emph{Virial expansion with Feynman diagrams\/}, Phys. Rev. A {\bf 84}, 053633 (2011).

\bibitem{Liu-2013}
X.-J. Liu, \emph{Virial expansion for a strongly correlated Fermi system and its application to ultracold atomic Fermi gases\/}, Phys. Rep. {\bf 524}, 37 (2013). 

\bibitem{Pini-2019} 
M. Pini, P. Pieri, and G.~C. Strinati, \emph{Fermi gas throughout the BCS-BEC crossover: Comparative study of $t$-matrix approaches with various degrees of self-consistency\/}, Phys. Rev. B {\bf 99}, 094502 (2019).

\bibitem{Ku-2012} 
M.~J.~H. Ku, A.~T. Sommer, L.~W. Cheuk, and M.~W. Zwierlein, \emph{Revealing the superfluid lambda transition in the universal thermodynamics of a unitary Fermi gas\/}, Science {\bf 335}, 563 (2012).

\bibitem{Zwerger-2016} 
W. Zwerger, Strongly interacting Fermi gases, in \emph{Quantum Matter at Ultralow Temperatures\/}, M. Inguscio, W. Ketterle, S. Stringari, and G. Roati (Eds.), Proceedings of the International School of Physics ``Enrico Fermi", 
vol. 191, pp. 63-142 (IOS Press, Amsterdam, 2016).

\bibitem{Carcy-2019} 
C. Carcy,  S. Hoinka, M.~G. Lingham, P. Dyke, C.~C.~N. Kuhn, H. Hu, and C.~J. Vale, \emph{Contact and sum rules in a near-uniform Fermi gas at unitarity\/}, Phys. Rev. Lett. {\bf 122}, 203401 (2019).

\bibitem{Jensen-2020} 
S. Jensen, C.~N. Gilbreth, and Y. Alhassid, \emph{Contact in the unitary Fermi gas across the superfluid phase transition\/}, Phys. Rev. Lett. {\bf 125}, 043402 (2020).

\bibitem{Rammelmueller-2021} 
L. Rammelm\"uller, Y. Hou,  J.~E. Drut, and J. Braun, \emph{Pairing and the spin susceptibility of the polarized unitary Fermi gas in the normal phase\/}, Phys. Rev. A {\bf 103}, 043330 (2021).

\bibitem{Schaefer-2021} 
T. Sch\"afer {\em et al.}, \emph{Tracking the footprints of spin fluctuations: a multimethod, multimessenger study of the two-dimensional Hubbard model\/}, Phys. Rev. X {\bf 11}, 011058 (2021).

\bibitem{Ma-1953}
S.~T. Ma, \emph{Interpretation of the Virtual level of the Deuteron\/}, Rev. Mod. Phys. {\bf 25}, 853 (1953).

\bibitem{Nussenzveig-1959}
H.~M Nussenzveig, \emph{The poles of the S-matrix of a rectangular potential well or barrier\/}, Nucl. Phys. {\bf 11}, 499 (1959).

\bibitem{Deltuva-2020}
A. Deltuva, M. Gattobigio, A. Kievsky, and M. Viviani,
\emph{Gaussian characterization of the unitary window for $N = 3$: Bound, scattering, and virtual states\/}, Phys. Rev. C {\bf 102}, 064001 (2020).

\bibitem{Marini-1998} 
M. Marini, F. Pistolesi, and G.~C. Strinati, \emph{Evolution from BCS superconductivity to Bose condensation: analytic results for the crossover in three dimensions\/}, Eur. Phys. J. B {\bf 1}, 151 (1998).

\bibitem{Pini-2020}
M. Pini, P. Pieri, M. J\"ager, J. Hecker Denschlag, and G.~C. Strinati, \emph{Pair correlations in the normal phase of an attractive Fermi gas\/}, New Jour. Phys. {\bf 22}, 083008 (2020).

\bibitem{Baym-1961}
G. Baym and L.~P. Kadanoff, \emph{Conservation laws and correlation functions\/}, Phys. Rev. {\bf 124}, 287 (1961).

\bibitem{Baym-1962}
G. Baym, \emph{Self-consistent approximations in many-body systems\/}, Phys. Rev. {\bf 127}, 1391 (1962).

\bibitem{Leyronas-2015}
M. Sun and X. Leyronas, \emph{High-temperature expansion for interacting fermions\/}, Phys. Rev. A {\bf 92}, 053611(2015).

\bibitem{Chin-2005} C. Chin and P.~S. Julienne,
\emph{Radio-frequency transitions on weakly bound ultracold molecules\/}, Phys. Rev. A {\bf 71}, 012713 (2005).

\bibitem{Smith-1989} H. Smith and H. H. Jensen, \emph{Transport Phenomena} (Oxford University Press, Oxford, 1989).

\bibitem{footnote-phasediagram} 
The transition from one color to another represents a gradual crossover between two physical regimes and its smoothness is rendered by mapping the color palette to sigmoid functions, 
that are centered at the boundaries defined for each sector in Sec.~\ref{sec:phdiag}.

\bibitem{Parish-2013} 
V. Ngampruetikorn, J. Levinsen, and M.~M. Parish, \emph{Pair correlations in the two-dimensional Fermi gas\/}, Phys. Rev. Lett. {\bf 111}, 265301 (2013).

\bibitem{Li-2023} 
X. Li et al., {\em Observation and quantification of pseudogap in unitary Fermi gases}, arXiv:2310.14024 (2023).

\bibitem{Combescot-2005} 
R. Combescot, \emph{Shift of the molecular bound state threshold in dense ultracold Fermi gases with Feshbach resonance\/}, New Jour. Phys. {\bf 5}, 86 (2003).

\bibitem{Dawson} 
 N. M. Temme, \emph{Error Functions, Dawson's and Fresnel Integrals\/}, in F.~W.~J. Olver, D.~M. Lozier,  R.~F.  Boisvert, C.~W. Clark  (eds.), NIST Handbook of Mathematical Functions, Cambridge University Press (2010).

\bibitem{Enss-2011} 
T. Enss, R. Haussmann, and W. Zwerger, \emph{Viscosity and scale invariance in the unitary Fermi gas\/}, Ann. Phys. {\bf 326}, 770 (2011).

\bibitem{Bruun-2007} 
G. M. Bruun and H. Smith, \emph{Shear viscosity and damping for a Fermi gas in the unitarity limit\/}, Phys. Rev. A {\bf 75}, 043612 (2007).

\bibitem{Massignan-2005} 
P. Massignan, G. M. Bruun, and H. Smith, \emph{Viscous relaxation and collective oscillations in a trapped Fermi gas near the unitarity limit\/}, Phys. Rev. A {\bf 71}, 033607 (2005).

\end{thebibliography}
\end{document}